%% file: iclr2027_conference.tex
\documentclass{article} % For LaTeX2e
\usepackage{iclr2027_conference,times}

\usepackage[utf8]{inputenc}
\usepackage[T1]{fontenc}
\usepackage{amsmath}
\usepackage{amssymb}
\usepackage{graphicx}
\usepackage{booktabs}
\usepackage{pifont}   % \ding for the capability marks
\usepackage{multirow}
\usepackage{microtype}
\usepackage{xcolor}
\usepackage{colortbl}   % \rowcolor for highlighted table rows
\usepackage{algorithm}
\usepackage{algpseudocode}
\usepackage[colorlinks=true]{hyperref}   % coloured link text instead of boxes

\usepackage{preprint-style}

\definecolor{venuegray}{gray}{0.55}          % venue tag after a method name
\definecolor{checkgreen}{RGB}{28,124,74}     % supported capability
\definecolor{crossred}{RGB}{186,54,48}       % absent capability
\definecolor{headertint}{RGB}{238,242,248}
\definecolor{eighighlight}{RGB}{219,229,242}
\definecolor{ladderbase}{RGB}{233,237,241}
\definecolor{laddermid}{RGB}{228,232,240}
\definecolor{linkurl}{RGB}{31,78,121}
\hypersetup{citecolor=checkgreen, linkcolor=crossred, urlcolor=linkurl}

\newcommand{\tablefont}{\fontsize{8}{9.2}\selectfont}
\newcommand{\tablefontdense}{\fontsize{7}{8.0}\selectfont}
\newcommand{\venue}[1]{\,\textcolor{venuegray}{\fontsize{5.8}{6.6}\selectfont [#1]}}
\newcommand{\cmark}{\textcolor{checkgreen}{\ding{51}}}
\newcommand{\xmark}{\textcolor{crossred}{\ding{55}}}

\title{Evolving Idea Graphs with Learnable Edits-and-Commits for Multi-Agent Scientific Ideation}

\author{%
  \textbf{Jiangwen Dong\textsuperscript{1}}\quad
  \textbf{Bo Li\textsuperscript{2}}\quad
  \textbf{Wanyu Lin\textsuperscript{1}*}\\[0.35em]
  \textsuperscript{1}The Hong Kong Polytechnic University\\
  \textsuperscript{2}The Hong Kong University of Science and Technology\\[0.3em]
  {\small\texttt{jiangwen.dong@connect.polyu.hk,\ bli@ust.hk,\ wan-yu.lin@polyu.edu.hk}}%
}

\iclrfinalcopy
\begin{document}

\maketitle

\begin{abstract}
LLM-empowered multi-agent systems offer new potential to accelerate scientific discovery by generating novel research ideas.
However, existing methods typically coordinate agents through temporary texts, such as drafts or chat logs; it is difficult to pinpoint the weaknesses in the generated ideas and how the agents refine them.
To this end, we introduce \textbf{Evolving Idea Graphs} (EIG), a graph-based multi-agent scientific ideation framework that can generate high-performance research ideas across various benchmark-native metrics, such as novelty, feasibility, and clarity.
Instead of coordinating solely through texts, EIG represents a partially formed proposal as an evolving idea graph, where nodes capture scientific claims and edges encode relations (e.g., support and conflict), enabling unresolved weaknesses to remain identifiable throughout the idea evolving process.
Specifically, a learned two-head controller operates over the evolving graph to guide the ideation: one head selects graph edits for agents to execute, while the other decides when the graph is ready for commit as final proposal synthesis.
On AI Idea Bench 2025 and LiveIdeaBench, EIG outperforms all compared systems on both automatic benchmark scores and blinded expert ratings.
Ablations show that the graph-based state, its typed relations, and the learned edit-and-commit control each contribute to final proposal quality.
Code: \href{https://github.com/Jiangwen-Dong/2026-Ideation}{\textit{Evolving-Idea-Graphs}}.
\end{abstract}

\input{1-introduction}
\input{2-related-work}
\input{3-method}
\input{4-experiment}
\input{5-conclusion}

\newpage
% Required by ICLR 2027. Does not count toward the page limit, but must not exceed one page.
\subsection*{AI use statement}
In this work, we used generative AI tools for language polishing of the manuscript and for reviewing and checking our code. We have not used generative AI tools for idea generation, research design, or the analysis and interpretation of results, and the remaining tasks that require disclosure are not applicable to this work. We have reviewed all AI-assisted work: the authors checked all edited text for correctness and originality, and every code change suggested during AI-assisted review was verified and tested against the reported results. We take responsibility for the final content of this work, including text, claims, and artifacts.

% Recommended by ICLR 2027. Does not count toward the page limit.
\subsection*{Reproducibility statement}
The method and its runtime are specified in Section~\ref{sec:method} and Appendix~\ref{app:implementation_details}, which give the node and edge schema, the role set and action space, candidate-slate construction, the four graph signals, the two-head critic architecture and its training configuration, and the calibration and runtime safeguards. Appendix~\ref{app:training_signals_split_hygiene} documents the offline supervision corpus, the group-level train/dev split, the held-out evaluation packet, and the inference-cost accounting, and Appendix~\ref{app:human_eval_protocol} gives the human-evaluation protocol and rubric. Code is available at the repository linked in the abstract.

% Recommended by ICLR 2027. Does not count toward the page limit.
\subsection*{Ethics statement}
The blinded expert study involved volunteer reviewers scoring anonymized machine-generated proposals, with no identifiable human-subjects data collected; Appendix~\ref{app:human_eval_pool} reports the review and consent procedure.

\bibliographystyle{iclr2027_conference}
\bibliography{Reference}

\appendix
\input{appendix}

\end{document}

%% file: 1-introduction.tex
\section{Introduction}

The rapid progress of scientific research makes automated exploration of new ideas increasingly valuable~\citep{park2023papers}, and LLM-based agentic systems have emerged as a promising paradigm for scientific ideation~\citep{wei2025agenticscience}. Recent methods retrieve literature and draft proposals from benchmark-defined inputs~\citep{wang2019paperrobot,xu2023exploring,gu2024scimuse,radensky2024scideator,wang2024scipip,su2025many}, and benchmarks such as AI Idea Bench 2025 and LiveIdeaBench provide released evaluators for the generated ideas~\citep{qiu2025aibench,ruan2026liveideabench,majumder2025discoverybench,jansen2024discoveryworld}. The remaining difficulty is that fluent proposals can be surface-plausible: their central claims, mechanisms, or validation plans may still contain internal contradictions or lack supporting evidence~\citep{si2025novelideas}. The challenge is therefore not only to produce proposal-shaped text, but to expose the internal formation of a scientific idea so that its consistency and coherence stay inspectable.

This is especially pronounced in multi-agent ideation~\citep{ueda2025dialogues,quan2026towards}, where role-specialized agents contribute complementary functions such as mechanism design, novelty checking, and evaluation planning. Specialization helps only if those partial contributions are coordinated coherently, yet most systems coordinate through drafts, dialogue histories, or reranked candidates~\citep{ueda2025dialogues,quan2026towards, wu2023autogen,li2023camel,chen2024agentverse,hong2024metagpt,qian2025scalingmac} (Table~\ref{tab:related_work_positioning}). Once unsupported claims, novelty conflicts, or missing evaluation components are absorbed into plain text, it becomes hard to see what is missing, who should intervene, and what to refine next.

\input{tables/related_work_positioning_table}

We therefore formulate scientific ideation as control over a persistent, structured state: a shared evolving graph whose nodes are proposal units such as claims, methods, and evaluation plans, and whose edges encode support, contradiction, and dependency. Such a state exposes internal structure rather than only fluent text: weak claims are localized in nodes, overall logic is adjusted through edges, and agents can read, edit, and finally commit to it. Two challenges follow. If several agents edit the same graph concurrently, their edits may interfere or become order-sensitive; and without a principled stopping criterion, the system may stop too early or over-refine. EIG therefore needs both a runtime that controls how parallel edits interact and a controller that decides which edits to apply and when the graph is ready for synthesis.

We introduce \textbf{Evolving Idea Graphs} (EIG), a graph-based multi-agent scientific ideation framework with two coupled components. {\em First, EIG maintains an evolving idea graph under a frozen-snapshot runtime}: in each iteration all active roles observe the same pre-round snapshot and propose edits independently, and only the selected edits are merged afterwards, which makes same-round proposals comparable alternatives rather than order-sensitive rewrites. {\em Second, EIG defines a shared-encoder two-head graph critic as the controller over that graph}: one head selects among validated role-local actions, while the other predicts whether the finalized post-round graph is ready for proposal synthesis. We ask whether graph-structured collaboration improves scientific ideation, and whether adaptive edit-and-commit decisions benefit under identical runtime constraints.

We evaluate EIG in a benchmark-faithful setting in which the system generates a structured proposal from benchmark-defined inputs without access to hidden target-paper fields or held-out annotations. On AI Idea Bench 2025 and LiveIdeaBench, EIG achieves the strongest benchmark results and the best blinded expert evaluation among the compared baselines, and ablations show that the graph-based state, its typed relations, and the learned edit-and-commit control each contribute to the final quality.

In summary, our paper makes these contributions:

\begin{itemize}
\item \textbf{The evolving idea graph as shared state.} EIG makes the evolving idea itself an explicit, persistent state for collaborative ideation, together with a frozen-snapshot parallel runtime that turns graph evolution into a clean edit-and-commit process. To our knowledge it is the first benchmark-faithful ideation framework to do so while keeping unresolved weaknesses visible during collaboration.
\item \textbf{Learned two-head control.} A shared-encoder graph critic selects validated role-local edits with one head and predicts commitment readiness of the merged graph with the other, matching the runtime split between local refinement and global stopping.
\item \textbf{Benchmark-faithful evaluation.} On AI Idea Bench 2025 and LiveIdeaBench, EIG outperforms all compared baselines on both automatic benchmark scores and blinded expert ratings, and ablations show that the graph-based state, its typed relations, and the learned edit-and-commit control each contribute to final quality.
\end{itemize}

%% file: tables/related_work_positioning_table.tex
\begin{table}[t]
\centering
\tablefont
\renewcommand{\arraystretch}{0.98}
\caption{Qualitative positioning of representative compared methods by high-level coordination attributes that distinguish their collaboration interfaces and control mechanisms. This is a component-level summary rather than an empirical comparison.}
\label{tab:related_work_positioning}
\setlength{\tabcolsep}{13.15pt}
\begin{tabular}{lccc}
\toprule
\rowcolor{headertint}
\textbf{Method} & \textbf{\shortstack[c]{Role\\ specialization}} & \textbf{\shortstack[c]{Persistent shared\\ editable state}} & \textbf{\shortstack[c]{Learned runtime\\ edit/stop control}} \\
\midrule
SciPIP\venue{arXiv'24}~\citep{wang2024scipip} & \xmark & \xmark & \xmark \\
VirSci\venue{ACL'25}~\citep{su2025many} & \cmark & \xmark & \xmark \\
AI-Researcher\venue{ICLR'25}~\citep{si2025novelideas} & \xmark & \xmark & \xmark \\
\rowcolor{eighighlight}
\textbf{EIG (Ours)} & \cmark & \cmark & \cmark \\
\bottomrule
\end{tabular}
\end{table}

%% file: 2-related-work.tex
\section{Related Work}

\paragraph{Scientific ideation and proposal-generation systems.}
Automated scientific ideation has evolved from early systems such as PaperRobot~\citep{wang2019paperrobot} and concept co-occurrence-based idea verbalization~\citep{xu2023exploring} to recent LLM-based systems such as SciMuse~\citep{gu2024scimuse}, SciDeator~\citep{radensky2024scideator}, SciPIP~\citep{wang2024scipip}, ResearchAgent~\citep{baek2025researchagent}, MotivGraph-SoIQ~\citep{lei2025motivgraphsoiq}, VirSci~\citep{su2025many}, and AI-Researcher~\citep{si2025novelideas}. These systems show that proposal generation benefits from stronger literature context, retrieval structure, critique-and-revision loops, or role specialization, while The AI Scientist~\citep{lu2024aiscientist} broadens the scope to autonomous research workflows. Human evaluations further show that novelty alone is not enough: feasibility, specificity, and technical grounding remain difficult for machine-generated ideas~\citep{si2025novelideas}. Our work stays in the proposal-generation setting, but shifts the focus from polished text generation to controlling the evolving intermediate state from which the final proposal is synthesized.

\paragraph{Text-mediated deliberation and multi-agent coordination.}
Recent reasoning systems improve performance by making generation iterative rather than single pass. ReAct~\citep{yao2023react}, Self-Refine~\citep{madaan2023selfrefine}, Reflexion~\citep{shinn2023reflexion}, Tree of Thoughts~\citep{yao2023treeofthoughts}, and LATS~\citep{zhou2024lats} all rely on textual feedback or search over textual trajectories. Multi-agent frameworks such as AutoGen~\citep{wu2023autogen}, CAMEL~\citep{li2023camel}, AgentVerse~\citep{chen2024agentverse}, ChatDev~\citep{qian2024chatdev}, MetaGPT~\citep{hong2024metagpt}, and MacNet~\citep{qian2025scalingmac}, together with the ideation-focused dialogue study of \citet{ueda2025dialogues}, further show the value of role specialization and communication protocols. Their working interface is nevertheless textual---messages, plans, critiques, or drafts---which hides which parts of an emerging idea remain unsupported or in conflict with prior work, and which evaluation components are still missing. EIG preserves the value of deliberation while turning the evolving idea itself into an explicit shared editable state rather than a transcript-centered process.

\paragraph{Structured state, memory, and graph-based control.}
Another line of work makes the intermediate state richer than a flat text buffer. Graph of Thoughts~\citep{besta2024graphthoughts}, Plan-on-Graph~\citep{chen2024planongraph}, Agent Planning with World Knowledge Model~\citep{qiao2024worldknowledgemodel}, S-DAG~\citep{dong2026sdag}, G-Memory~\citep{zhang2025gmemory}, and HybridFlow~\citep{dong2025hybridflow} use graph structure or hierarchical memory to support planning, decomposition, or experience reuse. In scientific ideation, ResearchAgent~\citep{baek2025researchagent} connects literature through an academic graph and MotivGraph-SoIQ~\citep{lei2025motivgraphsoiq} grounds ideation with a motivational knowledge graph, but these structures support retrieval or grounding rather than serving as the persistent collaborative object that agents edit round by round. Version-control-style agent frameworks such as AgentGit~\citep{li2025agentgit} and the Git Context Controller~\citep{wu2025gitcontextcontroller} instead apply commit, branch, and merge semantics to agent context or artifacts, and consistent reads of concurrently updated state are the classical snapshot-isolation problem~\citep{berenson1995critique}. EIG adopts those systems principles rather than claiming them: it versions a typed scientific idea graph, and its contribution is the learned policy for when to edit and when to commit. Table~\ref{tab:related_work_positioning} summarizes this component-level distinction.

%% file: 3-method.tex
\section{Method}
\label{sec:method}

\begin{figure}[t]
\centering
\includegraphics[width=\textwidth]{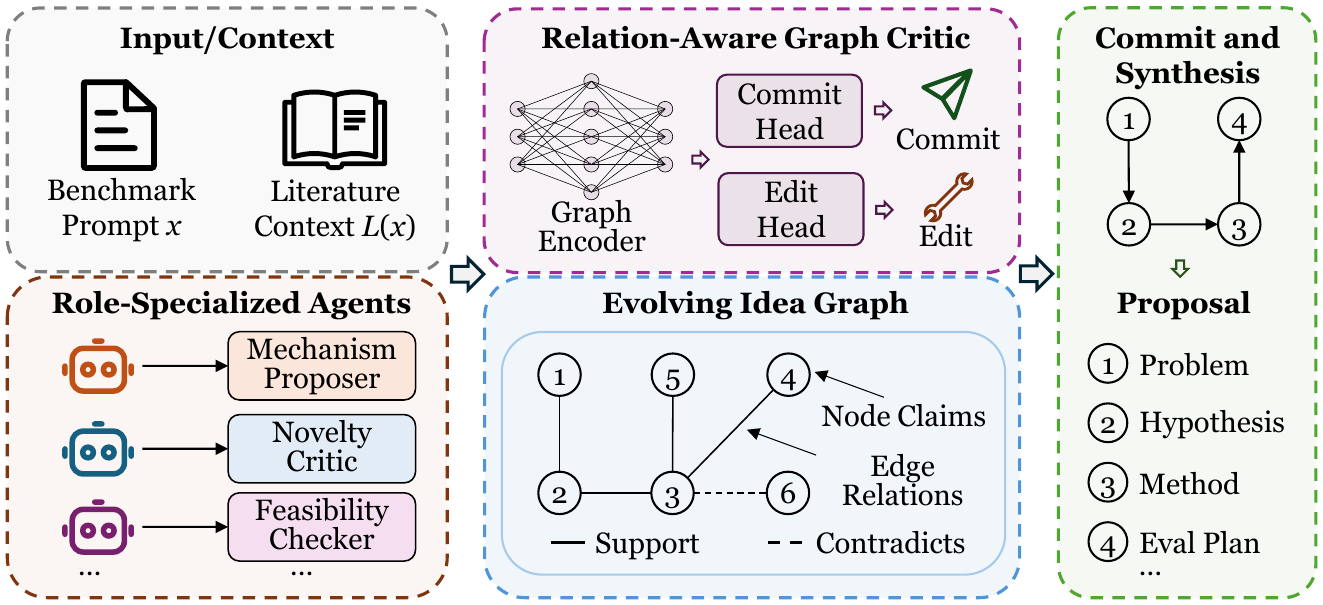}
\caption{Framework of EIG. Benchmark input and permitted literature context initialize role-specialized agents and an evolving idea graph. In each round, active roles propose role-local edits on a frozen graph snapshot; a shared graph encoder feeds an edit head for role-local action selection and a commit head for post-round stopping. When the updated graph is committed, the system synthesizes one structured research proposal.}
\label{fig:framework}
\end{figure}

Evolving Idea Graphs (EIG) treats multi-agent scientific ideation as control over a persistent structured state rather than repeated rewriting of one draft. A graph-based collaborative substrate is paired with a learned shared-encoder two-head graph critic that selects role-local edits and predicts when the realized post-round graph is ready for final proposal synthesis (Figure~\ref{fig:framework}).

\subsection{Problem Formulation}

Let $x$ be a benchmark-defined ideation input and $\mathcal{L}(x)$ the benchmark-permitted context. Every compared method must generate one structured proposal $y$ from the same benchmark-visible input under the same output contract: in AI Idea Bench 2025 only benchmark-safe topic and reference information is available, and in LiveIdeaBench only the keyword-conditioned prompt, never hidden target-paper fields or held-out scored idea text.

We assume role-specialized agents $\mathcal{M}=\{m_1,\ldots,m_K\}$, such as mechanism proposer, novelty examiner, feasibility critic, evaluation designer, and impact reframer. The challenge is not to make these roles contribute but to control how their partial contributions accumulate, since specialization helps only if it yields complementary structure rather than uncontrolled drift. EIG therefore separates two questions that text-only systems usually entangle: which local edit each role should make next, and when the shared state is coherent enough to stop and synthesize.

\subsection{Evolving Idea Graph as Collaborative State}

The first contribution of EIG is to make scientific ideation explicit as persistent structured collaborative state. At round $t$, the ideation state is a typed graph
\begin{equation}
G_t=(V_t,E_t),
\end{equation}
where nodes represent partial scientific units and edges their relations. The graph stores problems, hypotheses, methods, risks, novelty claims, evidence needs, and evaluation structure in a form that both agents and the graph critic can inspect. This persistent state lets roles add complementary structure without collapsing early to one draft, and keeps unsupported mechanisms, novelty overlap, and missing evaluation plans localized for repair. Appendix~\ref{app:schema} gives the node and edge vocabularies.

\subsection{Parallel Edit-and-Commit Control}

The second component of EIG is a parallel runtime that turns graph evolution into a well-defined control problem. Starting from the current graph $G_t$, a shared encoder produces a cached state representation
\begin{equation}
H_t = E_{\theta}(G_t,x).
\end{equation}
A lightweight activation gate selects the active roles $\mathcal{M}_t \subseteq \mathcal{M}$. All active roles then observe the same frozen snapshot $\bar G_t=G_t$ and independently propose validated role-local slates $\mathcal{S}_t^{(r)}$. Each slate draws on six schema-valid action types---\texttt{support}, \texttt{dependency}, \texttt{contradiction}, \texttt{evidence}, \texttt{repair}, and \texttt{skip}---so every candidate is a typed operation on the graph rather than free-form text. The action set is phased across rounds: the first round exposes missing support, dependency, and contradiction links, later rounds add evidence and repair, and commitment is withheld until round~3 (Appendix~\ref{app:role_scheduling}). Because every active role conditions on the same state, same-round alternatives stay comparable, within-round order effects shrink, and same-round proposals reflect complementary perspectives on one graph rather than an arbitrary update order.

For each active role, the edit head selects one decision
\begin{equation}
d_t^{(r)}=\arg\max_{a\in\mathcal{S}_t^{(r)}} q_{\theta}^{\mathrm{edit}}(H_t,r,a,x).
\end{equation}
A selected decision is a graph-critic output, not yet a graph update: each is materialized relative to the pre-round snapshot as a role-local patch $\Delta_t^{(r)}$, with \texttt{skip} mapping to an empty patch. Non-empty patches are then merged deterministically and applied to the shared graph,
\begin{equation}
G_{t+1}=T(G_t,\Delta_t),
\end{equation}
with $\Delta_t$ the realized merged patch set, that is, the deterministic post-selection realization of the already selected decisions. Separating selection from realization is what keeps role-local decisions comparable: they are conditioned on the same snapshot and cannot depend on within-round update order.

After the merge, the updated graph is encoded once more to obtain $H_{t+1}=E_{\theta}(G_{t+1},x)$, and the commit head predicts whether the realized post-round graph should stop:
\begin{equation}
p_{t+1}^{\mathrm{commit}} = q_{\theta}^{\mathrm{commit}}(H_{t+1},x).
\end{equation}
If the graph is not committed, the cached $H_{t+1}$ is reused in the next round. This post-round design matches the runtime semantics: edit selection is role-local and candidate-aware on $G_t$, whereas commitment is a graph-global judgment on the realized $G_{t+1}$, asking whether the merged graph already contains a stable problem--method--evaluation backbone worth synthesizing. Appendix~\ref{app:runtime_algorithm} gives the full algorithm and the precise relation between selected decisions, materialized patches, and commit rows.

\subsection{Learning the Graph Critic}

We train a shared-encoder two-head graph critic on the heuristic-curated label set collected under the same frozen-snapshot protocol. Its relation-aware encoder embeds node text, node and role types, and scalar state features over typed directed edges. The edit head scores each validated role-local candidate from graph, target-neighborhood, candidate, and state summaries, and the commit head scores the realized post-round graph from graph, state-text, and post-round summaries.

A protocol-matched heuristic teacher supplies those labels. Offline and under the same runtime, we profile 400 benchmark groups, build validated role-local candidate slates, label the teacher-selected action positive and every other candidate in the slate---including \texttt{skip}---negative, and give each merged post-round graph a binary continue-or-commit label. Because training and inference share the same frozen snapshots, candidate slates, merged graphs, and stopping decisions, the supervision is deployment-aligned. The critic is trained with a grouped slate-ranking loss for edit selection and a class-weighted binary cross-entropy for commit prediction on group-level train/dev splits.

In deployment the learned critic is the primary controller, with a lightweight post-hoc calibration driven by four graph signals; Appendices~\ref{app:candidate_slate}, \ref{app:two_head_graph_critic}, \ref{app:critic_calibration}, and \ref{app:training} give the slate construction, critic, calibration, safeguards, and training details. Profiling is a one-time offline cost: once trained, the critic adds no online model calls beyond the normal role and synthesis calls (Appendix~\ref{app:inference_cost_analysis}).

Once the graph is committed at round $\tau$, the system extracts the active problem--hypothesis--method--evaluation node set, with its supporting edges and attached evidence, as the proposal backbone, and synthesizes the final structured idea from that scaffold. The proposal therefore follows an explicit scientific backbone rather than the last conversation state alone.

%% file: 4-experiment.tex
\section{Experiments}

\subsection{Experiment Setup}

\paragraph{Datasets.}
We evaluate on AI Idea Bench 2025 (AIIB)~\citep{qiu2025aibench} and LiveIdeaBench~\citep{ruan2026liveideabench}. AIIB targets literature-grounded ideation and scores against hidden target-paper information; LiveIdeaBench checks robustness under lighter keyword-only context. We use a fixed subset of 512 held-out groups split evenly across the two benchmarks (256/256), and all claims are relative to it. The 512 groups are disjoint from the 400-group graph-critic corpus at the group level.

\paragraph{Baselines.}
We compare benchmark-faithful reproductions under a unified interface. \textbf{Direct} is a one-shot generator from the benchmark-visible packet; \textbf{Self-Refine}~\citep{madaan2023selfrefine} critiques and rewrites its own draft; \textbf{Graph of Thoughts}~\citep{besta2024graphthoughts} is a graph-based reasoning paradigm; \textbf{AI-Researcher}~\citep{si2025novelideas} is a multi-stage literature-grounded pipeline; \textbf{SciPIP}~\citep{wang2024scipip} is a structured planning-and-drafting pipeline; and \textbf{VirSci}~\citep{su2025many} is a discussion-oriented multi-agent proposal system. All receive the same inputs and emit the same schema, isolating control strategy from privileged access or output format.

\paragraph{Evaluation Metrics.}
We report the released benchmark score on each benchmark, rescaled to a common 0--10 range for readability, and a blinded expert evaluation with rubric scores and within-group rankings. Each benchmark keeps its native evaluator; the conversion is only a presentation-layer rescaling. The AIIB evaluator compares a generated idea against the hidden target paper's motivation and experimental design, including core-idea recovery against distractors, and the LiveIdeaBench evaluator scores originality, feasibility, substantive diversity, and cross-domain robustness. Runtime quantities such as rounds and critic overrides are used only for analysis, and Appendix~\ref{app:inference_cost_analysis} separates the one-time offline profiling cost.

\paragraph{Implementation Details.}
All EIG variants use the same parallel frozen-snapshot runtime and the same final proposal schema, instantiated with \textbf{Qwen3-8B}. The reproduced baselines use the same model family and input packet, keeping their own prompting logic and control flow. The controller is a shared-encoder two-head graph critic with \texttt{all-MiniLM-L6-v2} text embeddings, a two-layer relation-aware graph encoder (hidden size 128), and separate edit and commit heads. It is trained on grouped train/dev splits from a 400-group offline supervision corpus curated by a protocol-matched heuristic teacher, and uses lightweight graph-signal calibration at deployment (ablated in Appendix~\ref{app:critic_calibration}). The reproduced baselines are adapted from their public releases (Appendix~\ref{app:budget_matched}), and Appendix~\ref{app:implementation_details} gives the full configuration.

\subsection{Results}

Table~\ref{tab:main_results} reports automatic benchmark results on the fixed 512-group held-out subset, split evenly between AI Idea Bench 2025~\citep{qiu2025aibench} and LiveIdeaBench~\citep{ruan2026liveideabench}. Each row gives the released AIIB and LiveIdeaBench scores, their average, and traced generation tokens for cost context, and Table~\ref{tab:human_eval_results} reports the blinded expert evaluation. The three multi-stage baselines are budget-matched, so the comparison is not explained by a larger inference budget; error bars are standard deviations over three independent runs.

\input{tables/quality_main_table.tex}
\input{tables/human_eval_table.tex}

\paragraph{Automatic benchmark results.}
Table~\ref{tab:main_results} shows EIG reaching 7.69 on AI Idea Bench 2025 and 7.12 on LiveIdeaBench, for a 7.41 average: 0.46 above the budget-matched strongest baseline, AI-Researcher, and at least 0.53 above the others, although each receives at least EIG's traced inference budget. The ordering holds on both benchmarks, so the advantage is not an artifact of one released evaluator or visible-context regime, and it survives a generator swap to DeepSeek-V3.2 or GPT-4.1-mini (Appendix~\ref{app:backbone_sensitivity}).

\paragraph{Cost context.}
Figure~\ref{fig:budget_response} plots each method's pooled score against its traced token budget (Table~\ref{tab:main_results}). EIG reaches the highest pooled average at 13.75k tokens per sample, while the three budget-matched baselines spend 14.14k--16.42k and still trail by at least 0.46 points; scaling them further changes their pooled average by at most 0.02 points. Appendix~\ref{app:budget_matched} gives the protocol and per-run values.

\paragraph{Blinded expert evaluation.}
Table~\ref{tab:human_eval_results} reports blinded expert evaluation on 24 held-out groups balanced across the two benchmarks. EIG attains the best mean overall quality (4.01) and the best average within-group rank (1.64); the strongest baseline, VirSci, reaches 3.50 and 2.92. Paired against that baseline, overall quality improves by 0.51 (95\% CI [0.36, 0.65]) and average rank by 1.28 (95\% CI [0.88, 1.68]), both at $p<10^{-4}$. The largest gains are in significance and clarity, with smaller positive trends in novelty, feasibility, and context adherence. Because the expert rubric is independent of the released scorers, this agreement rules out a scorer-specific advantage, and Appendix~\ref{app:human_eval_uncertainty} shows the differences are stable across matched group-review blocks.

\input{tables/figure_pair.tex}
\subsection{Analysis}

We use ablations and trajectory analysis to answer three framework-centric questions: do graph-structured controller representations matter, does the learned controller use that state intentionally, and does learned edit-and-commit control adapt to episode difficulty? Table~\ref{tab:ablation_results} reports all of them on the same frozen 512-group subset, grouped by the module that varies: \textbf{EIG-Random} samples legal actions at random, \textbf{EIG-Heuristic} uses the signal heuristic, \textbf{EIG-NoEdit} keeps only learned commit control, and \textbf{EIG-NoCommit} only learned edit control; \textbf{EIG-Text} scores flattened text instead of the graph and \textbf{EIG-TextState} replaces the shared state and synthesis with a text record, while \textbf{EIG-UntypedEdges} erases the relation types. \textbf{EIG-NoCalibration} and the two \textbf{EIG-Sequential} rows form the calibration-and-runtime block.

\paragraph{Where the gains come from.}
The state-and-synthesis block of Table~\ref{tab:ablation_results} holds the text controller fixed: graph state and synthesis raise the pooled average from 6.43 to 6.68, erasing relation types while keeping nodes and connectivity costs 0.97, and swapping in the relation-aware controller adds 0.73 to reach the full-system 7.41, each delta at least 3.9$\times$ the pooled standard error of the mean. The controller ladder spans 0.73 points (6.68 to 7.41), while the two representation contrasts differ by only 0.01, so no single component accounts for the full 0.98-point gap between text-only coordination and EIG. A learned text controller also scores no better than random selection (the edits-and-commits block), so the gain lies in the structured action space and the typed claim logic, not in graph structure or controller scoring. Edit control outweighs commit control (0.47 against 0.32), and the policy beats its heuristic teacher by 0.52 points while changing 37.7\% of its edit decisions and 14.4\% of its commit decisions. Appendix~\ref{app:state_attribution} and Appendix~\ref{app:teacher_student_audit} give the per-benchmark contrasts.

\paragraph{Learned critic versus calibration.}
The \textbf{EIG-NoCalibration} row of Table~\ref{tab:ablation_results} separates the learned critic from the deployment calibration layer on the same held-out subset. Replacing the uncalibrated learned critic with the heuristic teacher \textbf{EIG-Heuristic} costs 0.39 points (6.89 against 7.28), and removing the calibration layer costs a further 0.13 (7.28 against the full-system 7.41), so the learned critic carries most of the controller gain. Calibration contributes a smaller but consistent margin on both benchmarks, 0.11 on AIIB and 0.14 on LiveIdeaBench, and it does so without adding parameters or changing the offline labels: it stabilizes borderline commitment decisions on immature graphs and boosts grounding-targeted edits rather than substituting for learned scoring. Appendix~\ref{app:critic_calibration} gives the four signals and the runtime safeguards.

\paragraph{Runtime mechanism.}
The sequential-update rows of Table~\ref{tab:ablation_results} replace the frozen-snapshot runtime: the pooled average falls from 7.41 to 6.97 under the learned controller and from 6.89 to 6.72 under the heuristic anchor, with every other design choice held fixed, on both benchmarks and across all three role orders. The runtime is therefore part of the measured gain, not a coordination convenience: the learned controller loses more when later roles observe earlier same-round mutations, but the frozen snapshot helps both controller families. Once later roles observe earlier same-round mutations, they no longer decide on a common state, and merging their contributions is order-sensitive again, which is what the frozen snapshot removes. Reading one shared snapshot also lets the role calls of a round run in parallel, so the same choice is the lower-latency execution mode. Appendix~\ref{app:frozen_snapshot_runtime} reports the role-order variation and the latency comparison.

\input{tables/ablation_table.tex}

\begin{figure}[!t]
\centering
\includegraphics[width=0.95\textwidth]{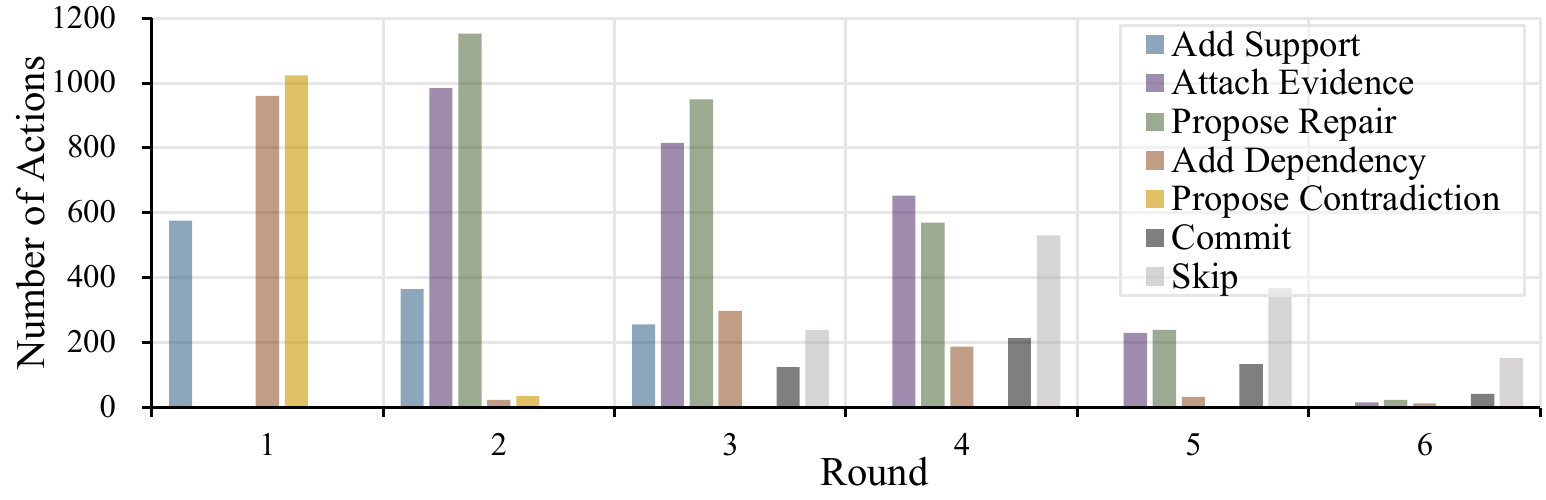}
\caption{Round-wise controller action distribution on the 512-group held-out EIG evaluation subset, normalized within each round; post-round commit decisions are shown separately.}
\label{fig:controller_behavior}
\end{figure}

\paragraph{Phased controller behavior.}
Selected actions are phased: dependency and contradiction edits account for 77.5\% of Round-1 role actions, evidence attachment and repair for 83.4\% of Round~2, and skip for 74.6\% by Round~6, while contradiction load falls from 0.97 to 0.00 and grounding rises from 0.30 to 0.68 (Table~\ref{tab:appendix_round_action_audit}, Figure~\ref{fig:controller_behavior}). The ordering is learned rather than imposed: with all six actions exposed in every round the shares barely move (73.8--75.0\% and 80.5--82.5\%), and the two unrestricted variants still score 6.96 and 7.01 against 7.41 (Table~\ref{tab:unrestricted_controller}). The controller therefore surfaces structural weaknesses before spending budget on grounding, then consolidates an existing scaffold.

\input{tables/unrestricted_controller_table.tex}

\paragraph{Adaptive stopping on hard cases.}
Commits cluster in the middle rounds---124 in Round~3, 214 in Round~4, 133 in Round~5, and 41 in Round~6---so the active set contracts from 512 episodes to 388, 174, and finally 41, retiring easier cases in favour of a hard-case tail. The head is not merely late: with all six actions exposed and commitment allowed from Round~1, only 6 of 512 episodes commit before Round~3, and the mean commit round stays at 4.16 against 4.18. Stopping therefore tracks graph maturity rather than a fixed round budget, so the later-round signal shifts in Figure~\ref{fig:round_signal} refine a shrinking tail. Appendix~\ref{app:unrestricted_controller} gives the stress-test protocol and the full traces.

\paragraph{Case evidence.}
Appendix~\ref{app:qualitative_case_study} traces one episode end to end: the edits move the modelled object from broad host--parasite interactions to life-cycle stages and host states, and replace generic robustness checks with temporal and host shifts. Appendix~\ref{app:case_literature} matches each change to independent literature. The episode is a hard case: the matched text-only draft is already fluent and plausible, so the gain is not stylistic. The graph localises the missing domain link, grounds it with evidence, and repairs it before commitment, turning a broad interaction model into a mechanism with named falsifiable tests.

%% file: tables/quality_main_table.tex
\begin{table*}[t]
\centering
\tablefont
\setlength{\tabcolsep}{12.9pt}
\renewcommand{\arraystretch}{0.98}
\caption{Automatic benchmark results on the fixed 512-group held-out subset. \textbf{AIIB} and \textbf{Live} are the released scores rescaled to 0--10, \textbf{Avg.} their mean, and \textbf{Total Tok.} traced generation tokens per sample. The three multi-stage baselines are budget-matched (Appendix~\ref{app:budget_matched}); error bars are three-run standard deviations.}
\label{tab:main_results}
\begin{tabular}{lcccc}
\toprule
\rowcolor{headertint}
\textbf{Method} & \textbf{AIIB} & \textbf{Live} & \textbf{Avg.} & \textbf{Total Tok.} \\
\midrule
Direct & 5.88{\scriptsize$\pm$0.11} & 6.56{\scriptsize$\pm$0.10} & 6.22 & 1.69k \\
Self-Refine\venue{NeurIPS'23}~\citep{madaan2023selfrefine} & 5.98{\scriptsize$\pm$0.10} & 6.62{\scriptsize$\pm$0.09} & 6.30 & 6.01k \\
Graph of Thoughts\venue{AAAI'24}~\citep{besta2024graphthoughts} & 6.24{\scriptsize$\pm$0.11} & 6.72{\scriptsize$\pm$0.10} & 6.48 & 17.79k \\
AI-Researcher\venue{ICLR'25}~\citep{si2025novelideas} & 7.10{\scriptsize$\pm$0.02} & 6.80{\scriptsize$\pm$0.02} & 6.95 & 14.81k \\
SciPIP\venue{arXiv'24}~\citep{wang2024scipip} & 6.94{\scriptsize$\pm$0.02} & 6.79{\scriptsize$\pm$0.03} & 6.87 & 16.42k \\
VirSci\venue{ACL'25}~\citep{su2025many} & 6.95{\scriptsize$\pm$0.02} & 6.81{\scriptsize$\pm$0.02} & 6.88 & 14.14k \\
\midrule
\rowcolor{eighighlight}
\textbf{EIG (Ours)} & \textbf{7.69}{\scriptsize$\pm$0.09} & \textbf{7.12}{\scriptsize$\pm$0.08} & \textbf{7.41} & 13.75k \\
\bottomrule
\end{tabular}
\end{table*}

%% file: tables/human_eval_table.tex
\begin{table*}[t]
\centering
\tablefont
\setlength{\tabcolsep}{4.1pt}
\renewcommand{\arraystretch}{0.98}
\caption{Blinded expert evaluation on 24 held-out groups, using the methods of Table~\ref{tab:main_results}. Each group is judged by three reviewers from a 12-person pool under anonymized identities, assigning 1--5 scores per aspect and within-group ranks; each cell is the mean over 72 judgments. Higher is better for rubric scores and lower for rank (Appendix~\ref{app:human_eval_uncertainty}).}
\label{tab:human_eval_results}
\begin{tabular}{lccccccc}
\toprule
\rowcolor{headertint}
\textbf{Method} & \textbf{Novelty} & \textbf{Signif.} & \textbf{Feas.} & \textbf{Clarity} & \textbf{Context} & \textbf{Overall} & \textbf{Avg. Rank} \\
\midrule
Direct & 2.18 & 2.72 & 2.76 & 2.71 & 2.00 & 2.64 & 5.74 \\
Self-Refine\venue{NeurIPS'23}~\citep{madaan2023selfrefine} & 2.90 & 2.79 & 3.03 & 3.07 & 3.06 & 2.79 & 5.51 \\
Graph of Thoughts\venue{AAAI'24}~\citep{besta2024graphthoughts} & 2.94 & 2.90 & 3.06 & 3.00 & 2.99 & 2.93 & 4.86 \\
AI-Researcher\venue{ICLR'25}~\citep{si2025novelideas} & 2.86 & 3.06 & 3.42 & 3.01 & 2.78 & 3.06 & 4.18 \\
SciPIP\venue{arXiv'24}~\citep{wang2024scipip} & 2.71 & 3.10 & 3.50 & 3.08 & 3.08 & 3.31 & 3.15 \\
VirSci\venue{ACL'25}~\citep{su2025many} & 2.90 & 3.14 & 3.71 & 3.28 & 2.99 & 3.50 & 2.92 \\
\midrule
\rowcolor{eighighlight}
\textbf{EIG (Ours)} & \textbf{3.04} & \textbf{3.78} & \textbf{3.88} & \textbf{3.74} & \textbf{3.15} & \textbf{4.01} & \textbf{1.64} \\
\bottomrule
\end{tabular}
\end{table*}

%% file: tables/figure_pair.tex
% The pooled-score and graph-signal figures are drawn as a matched pair: one
% float, two 0.48\textwidth minipages, each with its own caption and label, so
% they print side by side while keeping two consecutive figure numbers.  The two
% captions are deliberately kept to a similar length.
\begin{figure}[t]
\centering
\begin{minipage}[t]{0.48\textwidth}
\centering
\includegraphics[width=\linewidth]{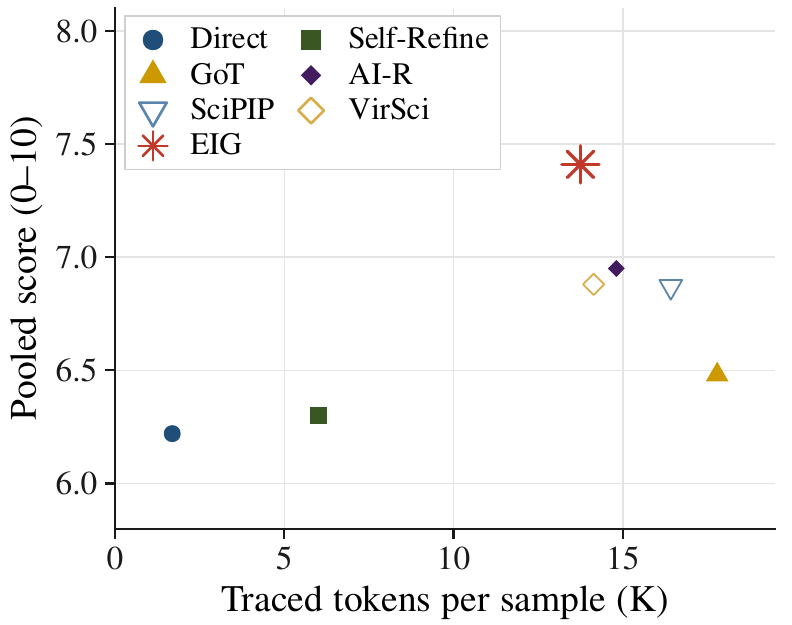}
\caption{Pooled score against the traced token budget for the methods of Table~\ref{tab:main_results}, at the configurations reported there (GoT = Graph of Thoughts, AI-R = AI-Researcher).}
\label{fig:budget_response}
\end{minipage}\hfill
\begin{minipage}[t]{0.48\textwidth}
\centering
\includegraphics[width=\linewidth]{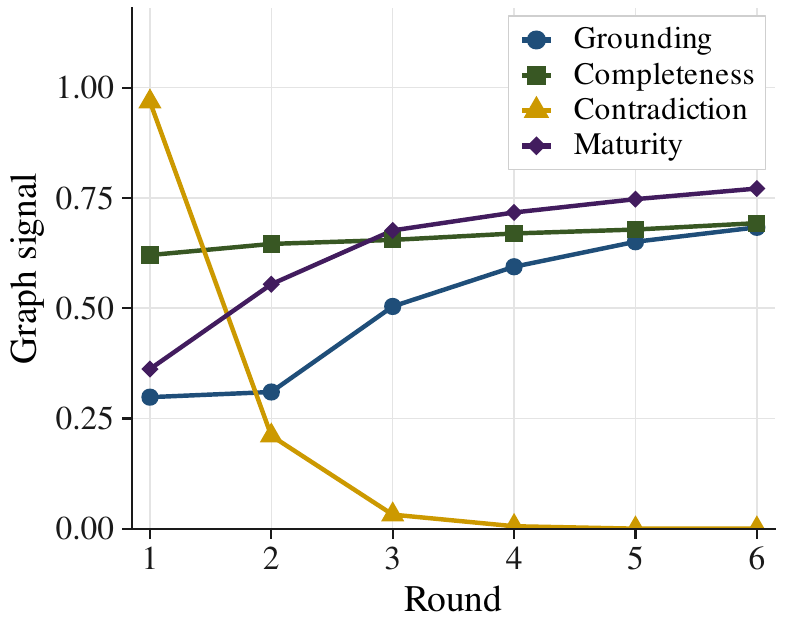}
\caption{Post-round graph-signal trajectories on the held-out EIG evaluation subset; maturity rises from 0.36 to 0.77 as the active set contracts to a hard-case tail of 41. Appendix~\ref{app:graph_utility_features} defines the four signals.}
\label{fig:round_signal}
\end{minipage}
\end{figure}

%% file: tables/ablation_table.tex
\begin{table}[t]
\centering
\tablefont
\renewcommand{\arraystretch}{0.94}
\caption{Ablations on the fixed 512-group held-out subset, grouped by axis, with the full system as reference. \textbf{State}/\textbf{Ctrl.}/\textbf{Synth.} are the shared state, control, and synthesis representations; \textbf{Edit}/\textbf{Commit} mark learned edit/commit control; \textbf{Calib.} marks the calibration layer (n/a for heuristic control). $\Delta$ is EIG (Ours) minus each variant; error bars are three-run, except the sequential rows (role orders).}
\label{tab:ablation_results}
% Four sub-tables, each with only the columns its rows need, padded to the text
% width so the stack reads as one block; the internal columns shift from block
% to block because each block carries only its own.
\setlength{\tabcolsep}{13.7pt}
\begin{tabular}{lcccccc}
\toprule
\rowcolor{headertint}
\textbf{Edits and commits} & \textbf{Edit} & \textbf{Commit} & \textbf{AIIB} & \textbf{Live} & \textbf{Avg.} & $\boldsymbol{\Delta}$ \\
\midrule
EIG-Random & \xmark & \xmark & 6.77{\scriptsize$\pm$0.10} & 6.60{\scriptsize$\pm$0.09} & 6.69 & +0.72 \\
EIG-Heuristic & \xmark & \xmark & 7.05{\scriptsize$\pm$0.08} & 6.72{\scriptsize$\pm$0.07} & 6.89 & +0.52 \\
EIG-NoEdit & \xmark & \cmark & 7.12{\scriptsize$\pm$0.07} & 6.76{\scriptsize$\pm$0.05} & 6.94 & +0.47 \\
EIG-NoCommit & \cmark & \xmark & 7.33{\scriptsize$\pm$0.10} & 6.85{\scriptsize$\pm$0.09} & 7.09 & +0.32 \\
\bottomrule
\end{tabular}

\vspace{1.0pt}
\setlength{\tabcolsep}{10.7pt}
\begin{tabular}{lccccccc}
\toprule
\rowcolor{headertint}
\textbf{State and synthesis} & \textbf{State} & \textbf{Ctrl.} & \textbf{Synth.} & \textbf{AIIB} & \textbf{Live} & \textbf{Avg.} & $\boldsymbol{\Delta}$ \\
\midrule
EIG-TextState & Text & Text & Text & 6.53{\scriptsize$\pm$0.03} & 6.33{\scriptsize$\pm$0.05} & 6.43 & +0.98 \\
EIG-Text & Graph & Text & Graph & 6.78{\scriptsize$\pm$0.09} & 6.58{\scriptsize$\pm$0.08} & 6.68 & +0.73 \\
EIG-UntypedEdges & Graph & Graph & Graph & 6.56{\scriptsize$\pm$0.03} & 6.32{\scriptsize$\pm$0.04} & 6.44 & +0.97 \\
\bottomrule
\end{tabular}

\vspace{1.0pt}
\setlength{\tabcolsep}{9.5pt}
\begin{tabular}{lccccccc}
\toprule
\rowcolor{headertint}
\textbf{Calibration and runtime} & \textbf{Calib.} & \textbf{Edit} & \textbf{Commit} & \textbf{AIIB} & \textbf{Live} & \textbf{Avg.} & $\boldsymbol{\Delta}$ \\
\midrule
EIG-NoCalibration & \xmark & \cmark & \cmark & 7.58{\scriptsize$\pm$0.09} & 6.98{\scriptsize$\pm$0.09} & 7.28 & +0.13 \\
EIG-Sequential-Heuristic & n/a & \xmark & \xmark & 6.80{\scriptsize$\pm$0.03} & 6.64{\scriptsize$\pm$0.05} & 6.72 & +0.69 \\
EIG-Sequential-DropIn & \cmark & \cmark & \cmark & 7.16{\scriptsize$\pm$0.05} & 6.78{\scriptsize$\pm$0.06} & 6.97 & +0.44 \\
\bottomrule
\end{tabular}

\vspace{1.0pt}
\setlength{\tabcolsep}{10.3pt}
\begin{tabular}{lccccccc}
\toprule
\rowcolor{headertint}
\textbf{Full system} & \textbf{State, Ctrl., Synth.} & \textbf{Edit} & \textbf{Commit} & \textbf{AIIB} & \textbf{Live} & \textbf{Avg.} & $\boldsymbol{\Delta}$ \\
\midrule
\rowcolor{eighighlight}
\textbf{EIG (Ours)} & Graph & \cmark & \cmark & \textbf{7.69}{\scriptsize$\pm$0.09} & \textbf{7.12}{\scriptsize$\pm$0.08} & \textbf{7.41} & -- \\
\bottomrule
\end{tabular}
\end{table}

%% file: tables/unrestricted_controller_table.tex
\begin{table}[t]
\centering
\tablefont
\setlength{\tabcolsep}{25.6pt}
\renewcommand{\arraystretch}{0.98}
\caption{Unrestricted-controller stress test on the held-out subset of Table~\ref{tab:ablation_results}. Both variants expose all six role-legal actions in every round and remove the phase mask, heuristic fallback, calibration, and safeguards; \textbf{Edits+Commit} also allows commitment from round~1.}
\label{tab:unrestricted_controller}
\begin{tabular}{lcc>{\columncolor{eighighlight}}c}
\toprule
\rowcolor{headertint}
\textbf{Metric} & \multicolumn{2}{c}{\textbf{Unrestricted}} & \textbf{EIG (Ours)} \\
\cline{2-3}
\rowcolor{headertint}
 & \textbf{Edits+Commit} & \textbf{Edits only} & \\
\midrule
AIIB & 7.13{\scriptsize$\pm$0.03} & 7.17{\scriptsize$\pm$0.03} & \textbf{7.69}{\scriptsize$\pm$0.09} \\
LiveIdeaBench & 6.79{\scriptsize$\pm$0.02} & 6.85{\scriptsize$\pm$0.03} & \textbf{7.12}{\scriptsize$\pm$0.08} \\
Avg. & 6.96 & 7.01 & \textbf{7.41} \\
\midrule
Round-1 structural & 73.8\% & 75.0\% & 77.5\% \\
Round-2 repair & 80.5\% & 82.5\% & 83.4\% \\
Commits in R1--R2 & 6/512 & 0/512 & 0/512 \\
Mean commit round & 4.16 & 4.22 & 4.18 \\
\bottomrule
\end{tabular}
\end{table}

%% file: 5-conclusion.tex
\section{Conclusion}

We presented Evolving Idea Graphs, a framework that treats scientific collaboration as control over persistent structured state rather than transient text. EIG couples an evolving idea graph with a learned two-head critic that selects role-local edits and predicts readiness for synthesis, and it achieves the strongest automatic benchmark results and the best blinded expert evaluation on AI Idea Bench 2025 and LiveIdeaBench. Making the intermediate scientific state explicit therefore appears useful for agentic systems that refine partially formed ideas rather than only generate polished text.

%% file: appendix.tex
\newpage

\section{Appendix Contents}
\label{app:contents}
\begingroup
\small
\setlength{\parindent}{0pt}
\setlength{\parskip}{0.18em}
\newcommand{\appendixsectionitem}[2]{%
  \par\addvspace{0.35em}%
  \textbf{\ref{#1}. #2}\nobreak\dotfill\pageref{#1}\par
}
\newcommand{\appendixsubsectionitem}[2]{%
  \hspace*{1.6em}\ref{#1}. #2\nobreak\dotfill\pageref{#1}\par
}
The appendix is organized as follows.

\appendixsectionitem{app:notation}{Notation}
\appendixsectionitem{app:parallel_runtime}{Full Parallel Runtime}
\appendixsectionitem{app:graph_schema}{Graph Schema and Runtime Artifacts}
\appendixsubsectionitem{app:graph_schema_nodes_edges}{Node and edge schema}
\appendixsubsectionitem{app:graph_schema_runtime_artifacts}{Selected decisions, patches, and commit rows}
\appendixsectionitem{app:implementation_details}{Concrete Implementation Details}
\appendixsubsectionitem{app:runtime_data_structures}{Runtime data structures}
\appendixsubsectionitem{app:role_scheduling}{Role scheduling and action space}
\appendixsubsectionitem{app:candidate_slate}{Candidate slate construction and heuristic teacher}
\appendixsubsectionitem{app:graph_utility_features}{Graph signals for label curation and calibration}
\appendixsubsectionitem{app:two_head_graph_critic}{Two-head graph critic}
\appendixsubsectionitem{app:licenses}{Licenses of existing assets}
\appendixsubsectionitem{app:critic_calibration}{Lightweight calibration and runtime safeguards}
\appendixsubsectionitem{app:state_attribution}{State, synthesis, and relation-typing ablations}
\appendixsubsectionitem{app:teacher_student_audit}{Teacher-student decision audit}
\appendixsectionitem{app:training_signals_split_hygiene}{Training Signals and Split Hygiene}
\appendixsubsectionitem{app:training_profiled_corpus}{Profiled corpus}
\appendixsubsectionitem{app:training_edit_supervision}{Edit supervision}
\appendixsubsectionitem{app:training_commit_supervision}{Commit supervision}
\appendixsubsectionitem{app:training_profiling_summary}{Profiling summary}
\appendixsubsectionitem{app:inference_cost_analysis}{Inference-time cost analysis}
\appendixsubsectionitem{app:budget_matched}{Matched inference-budget protocol}
\appendixsectionitem{app:controller_action_audit}{Round-Wise Controller Action Audit}
\appendixsubsectionitem{app:unrestricted_controller}{Unrestricted controller stress test}
\appendixsectionitem{app:frozen_snapshot_runtime}{Frozen-Snapshot vs Sequential-Update Runtime}
\appendixsectionitem{app:backbone_sensitivity}{Cross-Backbone Sanity Check With Selected Baselines}
\appendixsectionitem{app:human_eval_protocol}{Human Evaluation Protocol}
\appendixsubsectionitem{app:human_eval_packet}{Packet construction}
\appendixsubsectionitem{app:human_eval_pool}{Reviewer pool}
\appendixsubsectionitem{app:human_eval_blinding}{Assignment and blinding}
\appendixsubsectionitem{app:human_eval_packet_full}{Full instruction packet and compensation}
\appendixsubsectionitem{app:human_eval_rubric}{Scoring rubric}
\appendixsubsectionitem{app:human_eval_agreement}{Agreement summary}
\appendixsubsectionitem{app:human_eval_uncertainty}{Paired uncertainty}
\appendixsectionitem{app:qualitative_case_study}{Case Study}
\appendixsubsectionitem{app:case_background}{Case background}
\appendixsubsectionitem{app:case_trace}{Generated proposal and graph evolution}
\appendixsubsectionitem{app:case_interpretation}{Discussion}
\appendixsubsectionitem{app:case_literature}{Literature consistency of the induced direction}
\appendixsectionitem{app:limitations}{Limitations}
\endgroup

\section{Notation}
\label{app:notation}

Table~\ref{tab:notation} summarizes the main notation used in the parallel edit-and-commit formulation.

\input{tables/notation_table.tex}

\section{Full Parallel Runtime}
\label{app:parallel_runtime}
\label{app:runtime_algorithm}

Algorithm~\ref{alg:eig_parallel_appendix} gives the full forward path used by the parallel runtime.

\begin{algorithm}[ht]
\caption{Parallel EIG with edit-and-commit control}
\label{alg:eig_parallel_appendix}
\begin{algorithmic}[1]
\Require benchmark input $x$, permitted context $\mathcal{L}(x)$, roles $\mathcal{M}$, encoder $E_{\theta}$, edit head $q_{\theta}^{\mathrm{edit}}$, commit head $q_{\theta}^{\mathrm{commit}}$, calibration $\operatorname{Calibrate}$, round thresholds $\{\gamma_t\}$, maximum rounds $T_{\max}$
\Ensure final structured proposal $y$
\State Initialize typed idea graph $G_0 \leftarrow \operatorname{InitGraph}(x,\mathcal{L}(x))$
\State Encode once: $H_0 \leftarrow E_{\theta}(G_0,x)$
\For{$t = 0,\ldots,T_{\max}-1$}
    \State $\mathcal{M}_t \leftarrow \operatorname{ActivateRoles}(G_t,\mathcal{M})$
    \State Freeze snapshot $\bar G_t \leftarrow G_t$
    \State In parallel, collect role-local slates $\mathcal{S}_t^{(r)} \leftarrow \operatorname{Propose}(\bar G_t,x,r)$ for $r \in \mathcal{M}_t$
    \State Validate each $\mathcal{S}_t^{(r)}$ and add explicit \texttt{skip} if absent
    \Statex \Comment{The learned-critic path applies score calibration, diversity checks, and heuristic fallback (Appendix~\ref{app:critic_calibration}).}
    \ForAll{$r \in \mathcal{M}_t$}
        \State $d_t^{(r)} \leftarrow \arg\max_{a \in \mathcal{S}_t^{(r)}} q_{\theta}^{\mathrm{edit}}(H_t,r,a,x)$
    \EndFor
    \State Materialize each selected decision $d_t^{(r)}$ into a patch $\Delta_t^{(r)}$ relative to the pre-round snapshot; \texttt{skip} yields an empty patch
    \State $\Delta_t \leftarrow \{\Delta_t^{(r)} : r \in \mathcal{M}_t,\; \Delta_t^{(r)} \neq \emptyset\}$
    \State $G_{t+1} \leftarrow T(G_t,\Delta_t)$ \Comment{applied in deterministic role order}
    \State Encode once: $H_{t+1} \leftarrow E_{\theta}(G_{t+1},x)$
    \State $s_{t+1}^{\mathrm{commit}} \leftarrow q_{\theta}^{\mathrm{commit}}(H_{t+1},x)$
    \State $\tilde{s}_{t+1}^{\mathrm{commit}} \leftarrow \operatorname{Calibrate}(s_{t+1}^{\mathrm{commit}},G_{t+1})$ \Comment{lightweight signal calibration}
    \If{$\tilde{s}_{t+1}^{\mathrm{commit}} \ge \gamma_{t+1}$ and runtime guards pass}
        \State $\tau \leftarrow t+1$
        \State \Return $\operatorname{Synthesize}(G_{\tau},x,\mathcal{L}(x))$
    \EndIf
\EndFor
\State $\tau \leftarrow T_{\max}$
\State \Return $\operatorname{Synthesize}(G_{\tau},x,\mathcal{L}(x))$
\end{algorithmic}
\end{algorithm}

\section{Graph Schema and Runtime Artifacts}
\label{app:graph_schema}
\label{app:schema}

\subsection{Node and Edge Schema}
\label{app:graph_schema_nodes_edges}

\paragraph{Typed nodes.}
The idea graph uses a compact node schema:
\[
\mathcal{T}=
\left\{
\begin{array}{l}
\texttt{Problem},\ \texttt{Hypothesis},\ \texttt{Method},\ \texttt{Assumption},\\
\texttt{Risk},\ \texttt{EvalPlan},\ \texttt{NoveltyClaim},\ \texttt{EvidenceNeed},\ \texttt{Repair}
\end{array}
\right\}.
\]

\paragraph{Typed edges.}
Edges use a similarly compact relation schema:
\[
\mathcal{R}=
\left\{
\begin{array}{l}
\texttt{supports},\ \texttt{contradicts},\ \texttt{depends\_on},\ \texttt{overlaps\_prior},\\
\texttt{repairs},\ \texttt{refines},\ \texttt{requires\_evidence}
\end{array}
\right\}.
\]
This schema is intentionally lightweight. It is expressive enough to capture proposal structure, novelty conflicts, evidence needs, and repair actions, while remaining stable for multi-agent editing and replay collection.

\subsection{Selected Decisions, Patches, and Commit Rows}
\label{app:graph_schema_runtime_artifacts}

\paragraph{Runtime separation and deterministic realization.}
Parallel EIG distinguishes three runtime objects that would be conflated in a same-round sequential policy. First, the edit head chooses one validated role-local decision $d_t^{(r)}$ for each active role, including \texttt{skip}; these selected role decisions are the direct supervision targets for the edit head. Second, executing a selected role decision against the frozen graph snapshot yields a role-local patch $\Delta_t^{(r)}$. Deterministic merge here means deterministic realization of these already selected decisions: \texttt{skip} maps to an empty patch, candidate slates are deduplicated before selection by action kind, target identifiers, and payload, and the surviving selected actions are materialized into shared state under a fixed post-selection procedure. The key property is therefore order-freedom at decision time and determinism at realization time, rather than same-round dependence on arbitrary update order. Third, once the realized graph $G_{t+1}$ is obtained, the commit head produces a graph-level continue-or-commit judgment. This separation is what lets the two-head graph critic match the runtime semantics cleanly.

\paragraph{Concrete same-round example.}
Suppose two active roles observe the same frozen snapshot and both propose repairs around the same weak point in the graph. Their candidates are still scored against the same pre-round state, so edit selection remains comparable across roles. Only the selected role-local decisions are materialized afterward. In the heuristic-controller path, an additional diversity safeguard prefers a non-duplicative alternative when one is available; otherwise the runtime still follows the same fixed realization procedure. The purpose is not to solve arbitrary semantic disagreement between branches, but to make same-round execution reproducible and to avoid obvious redundant edits.

\paragraph{Merge order and conflict resolution.}
The deterministic merge materializes non-empty patches in a fixed action-kind order: \texttt{add\_contradiction\_edge}, \texttt{propose\_repair}, \texttt{add\_dependency\_edge}, \texttt{add\_support\_edge}, and \texttt{attach\_evidence}. This ordering reflects the intended round logic: flag unresolved conflicts first, then address them, then structure the backbone, then ground claims, and finally attach concrete evidence. Within the same kind, same-target conflicts are resolved by overwriting: a later patch in the fixed order replaces an earlier patch to the same node or edge. Cross-kind inconsistencies are reduced by the candidate-slate validator, which rejects candidates that would create direct structural contradictions---for example, a dependency edge paired with a contradiction edge between the same two nodes. Deduplication happens before selection, not during merge: the validator removes duplicate candidates by action kind, target identifiers, and payload, so the selected slate already contains at most one action per kind--target pair. When the heuristic teacher is used, the diversity safeguard mentioned above additionally prefers alternatives that target distinct nodes, but if no non-overlapping alternative is available, the runtime still proceeds with the fixed overwrite semantics. This procedure ensures reproducibility and eliminates obvious redundancy; it does not guarantee semantic optimality when role-local proposals genuinely disagree.

\section{Concrete Implementation Details}
\label{app:implementation_details}

\subsection{Runtime Data Structures}
\label{app:runtime_data_structures}

\paragraph{Objects.}
The implementation represents an idea graph as four persistent object types. Table~\ref{tab:runtime_objects} summarizes their key fields and runtime role. These fields are exported in the replay snapshots used for graph-critic training.

\paragraph{LLM backend.}
All paper-reported EIG runs use the same OpenAI-compatible backend configuration with \texttt{qwen3-8b} as the generator. The deployment configuration uses the DashScope-compatible API, temperature $0.2$, a default output budget of 1400 tokens for the internal EIG runtime, a 90-second timeout, and two retries. External reproduced baselines in the main comparison are likewise executed with the same \texttt{qwen3-8b} model family, while preserving method-specific interface settings such as their native output budgets when the reproduced implementation requires them.

\paragraph{Compute resources.}
All API-based inference uses the DashScope-compatible cloud endpoint for \texttt{qwen3-8b}. The graph critic is trained locally on a single NVIDIA GeForce RTX 3090 GPU (24~GB VRAM). Training the two-head critic for 8 epochs takes approximately 50 minutes. The full 512-group paper evaluation (including all baselines and three independent runs) requires approximately 20 hours of wall-clock time, including tracing and benchmark scoring.

\input{tables/runtime_objects_table.tex}

\subsection{Role Scheduling and Action Space}
\label{app:role_scheduling}

\paragraph{Roles.}
The deployed role set contains five persistent agents. These are \texttt{MechanismProposer}, \texttt{FeasibilityCritic}, \texttt{NoveltyExaminer}, \texttt{EvaluationDesigner}, and \texttt{ImpactReframer}. Each run instantiates one role-local branch for each of these five agents before the frozen-snapshot merge.

\paragraph{Round phases.}
The current control protocol uses a two-phase schedule. Round 1 is a structure-expansion round that prioritizes exposing missing support, dependency, or contradiction relations on the frozen graph snapshot. Round 2 and later rounds use repair-and-consolidation scheduling: the runtime continues to select one validated role-local action per active role, but now prioritizes actions that repair contradictions, reinforce weak grounding, or complete missing structural dependencies. In other words, the deployed appendix setting does not use a separate standalone ``stress-testing'' phase. The active-role gate activates all roles in the first round and then re-activates roles according to missing problem framing, weak grounding, incomplete dependencies, unresolved contradictions, or open feasibility work.

\paragraph{Validated actions.}
The active edit vocabulary consists of \texttt{add\_support\_edge}, \texttt{attach\_evidence}, \texttt{add\_dependency\_edge}, \texttt{add\_contradiction\_edge}, \texttt{propose\_repair}, and \texttt{skip}. For each active role, the runtime constructs a validated candidate slate and the edit policy selects exactly one role-local action, including \texttt{skip}. Only the validated selected action can materialize as a patch after deduplication and deterministic merge. The role-conditioned action set is intentionally compact: support and evidence actions improve graph grounding, dependency actions improve structural completeness, contradiction and repair actions address unresolved conflicts, and \texttt{skip} preserves mature states instead of forcing unnecessary edits.

\subsection{Candidate Slate Construction and Heuristic Teacher}
\label{app:candidate_slate}

\paragraph{Candidate expansion.}
For each active role, the system first constructs a heuristic baseline action and then expands it into a validated candidate slate over the deployed active edit vocabulary above. Candidate construction targets unsupported hypotheses, methods, novelty claims, and evaluation plans; unresolved contradictions; missing dependencies; and repairable nodes. Duplicate candidates are removed by action kind, target identifiers, and payload. Because the resulting slate size depends on the round phase, role, graph state, and deduplication outcome, we treat exact slate-count statistics as implementation details and do not headline them in the main paper.

\paragraph{Heuristic teacher and edit labels.}
Graph-critic training labels are curated from protocol-matched heuristic episodes. For each active role, the heuristic teacher simulates each valid candidate on the frozen graph, recomputes the four graph signals defined below, and selects the candidate that best reduces the dominant deficit while preserving overall maturity. Support and evidence actions target \texttt{grounding}, dependency actions target \texttt{completeness}, contradiction and repair actions target \texttt{contradiction\_load}, and \texttt{skip} is favored only when the current graph is already mature enough that further mutation is unlikely to help. The result is a within-slate weak preference signal: for a fixed frozen snapshot and validated candidate slate, one candidate is logged as positive and the remaining candidates in that same slate are negatives. During curation we also apply simple validity and low-gain safeguards so obviously degenerate edits and commits do not enter the training set.

\subsection{Graph Signals for Label Curation and Calibration}
\label{app:graph_utility_features}

\paragraph{Graph signals.}
The revised appendix presentation uses four graph signals. For round $t$, let $g_t$ denote \emph{grounding}, $r_t$ denote \emph{contradiction load}, $c_t$ denote \emph{completeness}, and $m_t$ denote \emph{maturity}. Grounding measures whether the current backbone claims are both structurally supported and explicitly evidenced. Contradiction load measures the remaining amount of unresolved conflict. Completeness measures whether the graph contains a usable scientific scaffold. Maturity is a post-round aggregate that increases when grounding and completeness are strong and contradiction load is low. Formally, the runtime computes
\[
\begin{aligned}
g_t
&=
\tfrac{1}{2}s_t^{\mathrm{sup}}
+\tfrac{1}{2}s_t^{\mathrm{evi}},\\
r_t
&=
0.65\,\ell_t^{\mathrm{edge}}
+0.35\,\ell_t^{\mathrm{open}},\\
c_t
&=
0.25\,q_t^{\mathrm{slot}}
+0.45\,q_t^{\mathrm{dep}}
+0.30\,q_t^{\mathrm{conn}},\\
m_t
&=
0.40\,g_t
+0.35\,c_t
+0.25\,(1-r_t).
\end{aligned}
\]
Here $s_t^{\mathrm{sup}}$ is support coverage over the grounding-focus nodes, $s_t^{\mathrm{evi}}$ is evidence coverage over the same focus set, $\ell_t^{\mathrm{edge}}$ is the unresolved-contradiction edge ratio, $\ell_t^{\mathrm{open}}$ is the ratio of contradiction targets that still lack repair, $q_t^{\mathrm{slot}}$ is claim-chain slot coverage, $q_t^{\mathrm{dep}}$ is dependency closure over the active hypothesis--method--evaluation backbone, and $q_t^{\mathrm{conn}}$ is structural connectivity of that backbone through support or dependency relations. All component terms are normalized to $[0,1]$ before aggregation.

\paragraph{Use in label curation and calibration.}
These coefficients should be read as normalized mixture weights over bounded graph-local statistics rather than benchmark-specific fit parameters. The same four-signal family is used in two places, but with different roles. Offline, the signals are used only to induce training labels inside a validated candidate slate and post-round commit trace. Online, the learned critic remains the primary scorer, and the signals are used only for a lightweight post-hoc calibration layer and runtime guards. In particular, the critic is trained from within-slate preference labels and binary post-round commit labels, not by direct regression to the four signal values themselves. The signals are computed entirely from the visible graph state and provide a stable controller-facing interface across the two benchmarks and different backbone LLM settings.

\paragraph{Coefficient design rationale.}
These coefficients were set by domain-driven design rather than empirical search. Grounding weights support coverage and evidence coverage equally ($\tfrac12$ each) because a well-grounded claim requires both structural support and explicit evidence; either alone is insufficient. Contradiction load weights unresolved edge conflicts higher ($0.65$) than open-target ratios ($0.35$) because an unresolved edge is a directly observable inconsistency, whereas an open target may simply be unaddressed yet. Completeness weights dependency closure highest ($0.45$) because a usable scientific scaffold requires causal links between problem, hypothesis, method, and evaluation; slot coverage ($0.25$) and structural connectivity ($0.30$) are secondary checks. Maturity balances the three readiness pillars: grounded claims ($0.40$), structural completeness ($0.35$), and resolved contradictions ($0.25$). All weights are simple fractions or near-fractions chosen once before experiments and never retuned.

\paragraph{Empirical robustness.}
The same fixed coefficients are used without modification across both benchmarks, all three generator backbones (Qwen3-8B, DeepSeek-V3.2, and GPT-4.1-mini; see Appendix~\ref{app:backbone_sensitivity}), and all reported ablations. This transfer is possible because the learned critic, not the coefficients, carries the main predictive burden: the coefficients shape the heuristic teacher and a lightweight calibration layer, but the \textbf{EIG-NoCalibration} row of Table~\ref{tab:ablation_results} shows that removing the calibration layer costs only 0.13 points, while replacing the full system with the heuristic teacher costs 0.52 points. Likewise, the cross-backbone sanity check (Table~\ref{tab:appendix_backbone_sanity}) shows that EIG retains the top ranking under both alternative backbones. These two results together confirm that the framework's performance is driven by learned graph structure and edit-and-commit control, not by brittle hand-tuned signal weights.

\paragraph{Graph maturity and empirical convergence.}
The maturity signal $m_t$ serves as a scalar summary of graph readiness. Because it aggregates grounding, completeness, and resolved contradictions, $m_t$ increases exactly when the graph is becoming more structurally sound. The commit head learns to score post-round graph states from the graph summary, state-text embedding, and compact post-round scalar features, and converts that score into a stop-or-continue decision. Because these inputs include structural and signal-derived summaries of graph readiness, the learned model effectively judges when the combined state indicates sufficient maturity for synthesis, rather than regressing to $m_t$ directly or waiting for a fixed formula to reach a preset value. This is empirical convergence monitoring: the signals provide reproducible, inspectable metrics of graph state, and the commit head generalizes from offline supervision to held-out deployment episodes. Figure~\ref{fig:round_signal} shows the empirical trajectory: contradiction load collapses early while grounding and maturity keep rising, which means the stopping decision is grounded in visible graph dynamics rather than in an arbitrary round budget. We do not claim a formal convergence guarantee to a unique fixed point; the framework instead provides reproducible graph-local metrics that let a learned model decide when further editing has diminishing returns.

\subsection{Two-Head Graph Critic}
\label{app:two_head_graph_critic}

\paragraph{Encoder inputs.}
The graph critic uses a shared relation-aware graph encoder. Each node input concatenates a text embedding, node-type embedding, role embedding, confidence scalar, and evidence-count scalar. Two relation-specific message-passing layers propagate information along typed directed edges, using the edge relation and resolution flag.

\paragraph{Edit head.}
The edit head scores one candidate at a time from the graph summary, target-node summary, local-neighborhood summary, candidate text embedding, candidate-kind embedding, and flattened state-text embedding.

\paragraph{Commit head.}
The commit head scores a post-round graph from the graph summary, state-text embedding, and three compact post-round scalar features: support coverage, unresolved contradiction ratio, and graph maturity.

\paragraph{Encoder architecture.}
The relation-aware encoder projects each node input into a $d$-dimensional hidden vector ($d = 128$). The node input concatenates a 384-dimensional sentence-transformer text embedding, a 16-dimensional node-type embedding, a 16-dimensional role embedding, a confidence scalar, and an evidence-count scalar. A two-layer MLP with ReLU maps this concatenated input to $h_v^{(0)} \in \mathbb{R}^d$, and invalid (padding) nodes are zeroed out by the graph mask before message passing begins and after each layer.

Each of the two message-passing layers proceeds as follows. For a directed edge $(u \to v)$ of relation type $r$, the outgoing message is $m_{u \to v}^{(r)} = W_r^{(l)} h_u^{(l)}$, where $W_r^{(l)} \in \mathbb{R}^{d \times d}$ is a relation-specific linear transform (no bias). The message is weighted by an edge importance factor $w_{u \to v} = \mathbb{1}[\text{edge active}] \cdot (1 + 0.1 \cdot \text{resolved}_{u \to v})$, so resolved edges receive a 10\% stronger signal. Aggregated messages at $v$ are computed by scatter-addition over all incoming edges:
\begin{equation}
\tilde{m}_v^{(l)} = \sum_{u : (u,v) \in E} w_{u \to v} \cdot m_{u \to v}^{(r_{u,v})} .
\end{equation}

Each node also accumulates three local structural statistics by the same scatter-addition pattern: in-degree ($s_v^{\text{in}}$), resolved in-degree ($s_v^{\text{resolved}}$), and out-degree ($s_v^{\text{out}}$). The update MLP consumes the concatenation $[h_v^{(l)}; \tilde{m}_v^{(l)}; s_v^{\text{in}}; s_v^{\text{resolved}}; s_v^{\text{out}}] \in \mathbb{R}^{2d+3}$, applies a two-layer MLP with ReLU, and the result is added residually and layer-normalized:
\begin{equation}
h_v^{(l+1)} = \text{LayerNorm}\bigl(h_v^{(l)} + \text{MLP}_{\text{update}}^{(l)}([h_v^{(l)}; \tilde{m}_v^{(l)}; s_v^{\text{in}}; s_v^{\text{resolved}}; s_v^{\text{out}}])\bigr) .
\end{equation}

\paragraph{Pooling and head computation.}
After the second message-passing layer, node states $h_v^{(2)}$ are pooled into summaries by masked mean aggregation. Let $V_{\text{mask}} \subseteq V$ denote the valid nodes for a given graph (or sub-mask for target / neighborhood). The pooled summary is
\begin{equation}
\bar{h} = \frac{1}{|V_{\text{mask}}|} \sum_{v \in V_{\text{mask}}} h_v^{(2)} .
\end{equation}
The edit head receives the concatenation $[\bar{h}_{\text{graph}}; \bar{h}_{\text{target}}; \bar{h}_{\text{neighbor}}; e_{\text{text}}^{\text{cand}}; e_{\text{kind}}^{\text{cand}}; e_{\text{text}}^{\text{state}}]$, where $e_{\text{text}}^{\text{cand}}$ and $e_{\text{text}}^{\text{state}}$ are sentence-transformer embeddings and $e_{\text{kind}}^{\text{cand}}$ is a 16-dimensional candidate-kind embedding. The edit head is a two-layer MLP with ReLU and scalar output. The commit head receives $[\bar{h}_{\text{graph}}; e_{\text{text}}^{\text{state}}; f_{\text{graph}}]$, where $f_{\text{graph}} \in \mathbb{R}^3$ is the vector of three post-round scalar features, and similarly applies a two-layer MLP with ReLU to produce a scalar commit score.

\paragraph{Training configuration.}
The deployed controller is the shared-encoder two-head graph critic used for the paper evaluations. Text inputs are embedded with the sentence-transformer \texttt{all-MiniLM-L6-v2}, which yields 384-dimensional text features. The relation-aware graph encoder uses hidden size 128, and training uses AdamW with learning rate $10^{-3}$, batch size 16, and 8 epochs. We optimize grouped slate-ranking loss for edit selection and class-weighted binary cross entropy for commit prediction; the binary commit loss uses a positive-class weight of 1.115 to compensate for label imbalance.

\subsection{Licenses of Existing Assets}
\label{app:licenses}

The benchmarks, models, and tools used in this paper are credited below with their licenses:

\begin{itemize}
    \item \href{https://github.com/yansheng-qiu/AI_Idea_Bench_2025}{\textbf{AI Idea Bench 2025}}~\citep{qiu2025aibench}: Apache-2.0.
    \item \href{https://github.com/x66ccff/liveideabench}{\textbf{LiveIdeaBench}}~\citep{ruan2026liveideabench}: MIT License.
    \item \href{https://huggingface.co/Qwen/Qwen3-8B}{\textbf{Qwen3-8B}}~\citep{yang2025qwen3}: Apache-2.0.
    \item \href{https://huggingface.co/sentence-transformers/all-MiniLM-L6-v2}{\textbf{all-MiniLM-L6-v2}}~\citep{reimers2019sentencebert}: Apache-2.0.
\end{itemize}

\subsection{Lightweight Calibration and Runtime Safeguards}
\label{app:critic_calibration}

\paragraph{Calibration.}
The deployed controller uses the learned two-head critic as the primary scorer. We then apply a small bounded graph-signal calibration based on \texttt{grounding}, \texttt{contradiction\_load}, \texttt{completeness}, and \texttt{maturity} so action selection and commitment are more intentional on obviously immature or already mature states. This refinement is lightweight: it does not change the offline labels or the critic weights, it does not replace the critic's candidate ranking with a heuristic policy, and it acts only as a post-hoc bias layer on top of the learned scores unless a runtime safeguard explicitly triggers fallback.

Concretely, calibration operates as a bounded additive bias on the learned scores. For edit selection, if grounding is low and contradiction load is high, the calibration layer adds a small positive bias to candidates that target grounding or contradiction repair; conversely, if maturity is high and contradiction load is low, it adds a small negative bias to non-skip actions, making the controller more selective. For commit decisions, if maturity is high and the learned commit score is borderline, calibration can lower the commit threshold; if grounding is low or contradiction load is high, it raises the threshold to prevent premature stopping. The bias magnitudes are capped at a small fraction of the learned score range so the critic's ranking is preserved unless the graph state is in an extreme regime. These bounds are set once and are not tuned per-benchmark.

\paragraph{Runtime safeguards.}
We additionally use simple runtime safeguards to prevent low-value or invalid overrides. The controller falls back to the heuristic action when runtime vocabulary tokens are unmapped, required target nodes are missing, or a low-gain kind swap would replace a plausible heuristic action. Edit overrides and commit decisions are also thresholded by calibrated margins and confidence. These are implementation stabilizers rather than the main learning contribution.

\paragraph{What calibration adds.}
The \textbf{EIG-NoCalibration} row of the main-text ablation (Table~\ref{tab:ablation_results}) separates a learned-critic component from a calibration component, and the learned two-head critic is the dominant driver of quality: it accounts for the larger part of the gain, with the lightweight calibration layer contributing the remainder (6.89, 7.28, and 7.41 for the heuristic teacher, the uncalibrated critic, and the full system). The calibration benefit is slightly larger on LiveIdeaBench (+0.14) than on AI Idea Bench 2025 (+0.11), suggesting that the sparser context regime benefits more from the additional stability that calibration provides.

This pattern supports the paper's architectural choice to treat the learned critic as the primary controller and calibration as a safety layer rather than as a replacement for learning. Calibration improves trajectory stability: it prevents premature commit on episodes whose contradiction load is still high, and it boosts grounding-targeted edits when the scaffold is immature. These are exactly the edge cases where a pure learned score, trained on average-case supervision, can be under-confident or over-confident, and correcting them costs far less than the learned critic itself. Appendix~\ref{app:graph_utility_features} defines the four graph signals and their mixture weights; the same signal family is used for heuristic label curation and for deployment calibration, which keeps the offline and online interfaces consistent.

\subsection{State, Synthesis, and Relation-Typing Ablations}
\label{app:state_attribution}

\paragraph{Matched text-state variant.}
EIG-TextState replaces the shared collaborative state with a structured text record and synthesizes the final proposal from that record, while keeping the role set, generator, runtime schedule, number of calls, inference budget, and output schema unchanged. Because a text state cannot be consumed by the relation-aware graph controller, EIG-TextState uses the same text controller as EIG-Text, and EIG-Text keeps graph state and graph synthesis with that same text controller. The two variants therefore differ only in the collaborative state and the synthesis backbone, with controller architecture, weights, and capacity held fixed.

\paragraph{Relation-typing variant.}
EIG-UntypedEdges keeps graph state, graph synthesis, and the learned controller, but replaces every typed relation with a generic \texttt{related\_to} edge. The node set, node types, connectivity, and action vocabulary are unchanged, so the comparison isolates the information carried by relation types rather than the presence of graph structure.

\paragraph{Results.}
The state-and-synthesis block of Table~\ref{tab:ablation_results} reports both contrasts on the same 512-group held-out subset. With the text controller fixed, graph state and graph synthesis raise the pooled average from 6.43 to 6.68, and the same increase appears on AI Idea Bench 2025 (6.53 to 6.78) and on LiveIdeaBench (6.33 to 6.58). The complementary contrast then holds graph state and synthesis fixed and reaches the full-system 7.41 through the relation-aware controller. Erasing relation types while preserving nodes and connectivity instead reduces the average from 7.41 to 6.44. The first contrast isolates the combined state-and-synthesis package and the second isolates relation typing; neither further separates persistent storage from final synthesis, so we do not claim universal graph superiority.

\subsection{Teacher-Student Decision Audit}
\label{app:teacher_student_audit}

\paragraph{Setup.}
To test whether the learned critic mainly imitates its heuristic teacher, we re-score the submitted signal heuristic counterfactually on the exact logged learned graphs and candidate slates, using a separate teacher diversity history and the submitted commit thresholds on the exact learned post-round graph. Both policies therefore decide on identical states and differ only in the decision rule. The audited EIG decision is the final submitted choice, after learned scoring, graph-signal calibration, diversity checks, heuristic fallback, and runtime safeguards.

\paragraph{Quality.}
Table~\ref{tab:teacher_student} reports the paired result. On identical states, EIG improves over the heuristic by 0.64 on AI Idea Bench 2025, 0.40 on LiveIdeaBench, and 0.52 on the pooled average, with paired intervals that exclude zero on both benchmarks. Read together with the calibration decomposition in Table~\ref{tab:ablation_results}---heuristic 6.89, learned critic without calibration 7.28, full EIG 7.41---the improvement does not come only from the heuristic signal family.

\paragraph{Decision agreement.}
Table~\ref{tab:teacher_student_agreement} shows how the improvement is distributed. The final policy changes 2,552 of 6,768 local edit decisions (37.7\%) while agreeing on 85.6\% of post-round commit decisions, so the learned gain comes mainly from edit selection and the stopping policy is largely preserved. Teacher dependence remains a limitation: the heuristic is part of the training specification, and we do not claim invariance to a substantially different supervision policy.

\input{tables/appendix_teacher_student_table.tex}

\input{tables/appendix_teacher_student_agreement_table.tex}

\section{Training Signals and Split Hygiene}
\label{app:training_signals_split_hygiene}
\label{app:training}

\subsection{Profiled Corpus}
\label{app:training_profiled_corpus}

\paragraph{Group-level collection.}
The two-head graph critic is supervised from a bounded offline supervision-labeling corpus collected under the same frozen-snapshot parallel protocol used at deployment. We profile 400 benchmark groups, balanced as 200 AI Idea Bench 2025 groups and 200 LiveIdeaBench groups.

\paragraph{Split hygiene.}
The graph-critic train/dev split is performed by group rather than by row: 300 groups are used for training and 100 for development, with each benchmark contributing 150 training groups and 50 development groups. This group-level split is important because one profiled run can yield many correlated edit candidates and commit states from the same underlying problem.

\paragraph{Held-out paper-eval packet.}
The final paper evaluation uses a separate frozen 512-group packet, balanced as 256 AI Idea Bench 2025 groups and 256 LiveIdeaBench groups. We keep this packet fixed across all compared methods and across repeated runs so that differences reflect method behavior under the same benchmark-visible inputs rather than variation in evaluation instances. This balance across the two benchmarks is intended to reduce obvious single-benchmark skew, but we do not treat the packet as an exhaustive census of the full benchmark populations. The split-hygiene audit reports zero benchmark-group overlap between this 512-group paper-eval packet and the 400-group graph-critic train/dev corpus.

\subsection{Edit Supervision}
\label{app:training_edit_supervision}

\paragraph{Filtering and counts.}
Starting from the curated heuristic episodes above, we retain only states with a valid snapshot reference, parseable candidate kind and target fields, and exactly one teacher-selected candidate. This produces 5,327 pre-action edit states and 63,833 candidate rows, split into 48,004 training candidate examples and 15,829 development candidate examples.

\subsection{Commit Supervision}
\label{app:training_commit_supervision}

\paragraph{Filtering and counts.}
For commit supervision, we package the realized graph after deterministic merge rather than the pre-round graph. Each retained post-round state carries the heuristic teacher's binary continue-or-commit label. We retain only states with a valid merged graph, a parseable round record, and a commit label in the expected schema. The resulting commit corpus contains 1,725 post-round states, with 400 commit positives and 1,325 continue examples; 1,294 states are used for training and 431 for development.

\subsection{Profiling Summary}
\label{app:training_profiling_summary}

\input{tables/training_signal_profile_table.tex}

\paragraph{Cost interpretation.}
Table~\ref{tab:training_signal_profile} reports the profiling overhead, which scales linearly with the number of profiled groups and the maximum number of rounds, while candidate-slate construction turns each profiled run into many supervised edit examples. The overhead is therefore a one-time offline profiling cost: once the graph critic is trained, edit and commit decisions are made by the learned model and do not add extra online language-model calls beyond the normal role proposal and final proposal synthesis calls. The held-out paper-eval split is constructed after excluding all graph-critic train/dev groups, and the disjointness audit reports zero overlap.

\subsection{Inference-Time Cost Analysis}
\label{app:inference_cost_analysis}

\input{tables/qwen_v2_inference_cost_table.tex}

\paragraph{Online cost profile.}
Table~\ref{tab:qwen_v2_inference_cost} reports traced inference-time generation tokens from the available Qwen3-8B paper-evaluation batch summaries. EIG uses 13.75k traced tokens per selected run, compared with 17.79k for Graph of Thoughts, 11.12k for AI-Researcher, 6.14k for VirSci, 6.01k for Self-Refine, 6.51k for SciPIP, and 1.69k for Direct. Relative to Direct, EIG uses 8.16$\times$ more traced tokens, but it is below Graph of Thoughts and only 1.24$\times$ above AI-Researcher. These traces count online generation calls only; benchmark scoring calls and the one-time offline graph-critic profiling corpus are excluded.

\paragraph{Interpretation.}
The additional online cost is driven more by prompt-side state than by completion length. EIG uses more prompt tokens than AI-Researcher (12.23k versus 8.91k), which is consistent with carrying persistent graph state, role-local candidate context, and merged-graph summaries across rounds. Its completion-token count remains lower than Graph of Thoughts and AI-Researcher. The one-time offline profiling cost used to train the graph critic remains separate and is reported in Table~\ref{tab:training_signal_profile}.

\subsection{Matched Inference-Budget Protocol}
\label{app:budget_matched}

\paragraph{Motivation.}
EIG carries a persistent graph across rounds, so its traced inference budget exceeds that of the single-pass and shallow multi-stage baselines. To separate quality from budget, every baseline whose public source code we can instantiate is additionally run at a budget that meets or exceeds EIG's, and the budget-matched runs are used for the main comparison in Table~\ref{tab:main_results}.

\paragraph{Scaling procedure.}
Each adapter is built from the baseline's public source code and receives additional method-native inference rather than padded output. AI-Researcher generates five seeds, filters and expands four, then ranks; SciPIP integrates and separately expands two ideas before synthesis; VirSci runs six role-cycled discussion turns before team synthesis. All methods use Qwen3-8B, identical benchmark-visible inputs and output schema, and the same 512-group held-out subset with three independent runs.

\paragraph{Token accounting and result.}
Table~\ref{tab:matched_budget} reports the as-submitted and budget-matched rows for the three scalable baselines, with EIG as the unscaled reference.

Tokens are average traced online generation tokens per held-out sample under the submitted aggregation; benchmark scoring calls and the one-time offline graph-critic profiling corpus are excluded. The three scaled baselines receive 14.81k, 16.42k, and 14.14k tokens, or 1.08$\times$, 1.19$\times$, and 1.03$\times$ EIG's 13.75k, so none of them is disadvantaged on budget; Graph of Thoughts already exceeds EIG's budget at 17.79k without modification. Scaling raises each baseline to at or above its as-submitted pooled average, and EIG still leads the strongest of them by 0.46 pooled points. Direct and Self-Refine are single-pass generators with no comparable scaling path. The three adapters build on the public releases of AI-Researcher (\texttt{NoviScl}), SciPIP (\texttt{cheerss}), and Virtual-Scientists (\texttt{open-sciencelab}), preserving each source configuration and changing only the method-native inference budget.

\input{tables/appendix_matched_budget_table.tex}

\section{Round-Wise Controller Action Audit}
\label{app:controller_action_audit}

\paragraph{Purpose.}
The main paper uses aggregate trajectory evidence to argue that EIG first exposes structural weaknesses, then repairs and grounds them, and finally commits only when the graph is mature enough for synthesis. Table~\ref{tab:appendix_round_action_audit} provides the exact round-wise controller trace behind that claim on the same 512-group held-out subset used in the main paper.

\input{tables/appendix_round_action_audit_table.tex}

\paragraph{Interpretation.}
Two patterns are most important. First, the early rounds are visibly phased: Round~1 is dominated by dependency and contradiction edits, whereas Rounds~2 and~3 shift toward evidence attachment and repair. This is the concrete behavioral signature of the framework's intended use of explicit graph state: the controller uses the graph to surface unresolved weaknesses before spending later rounds on targeted repair and grounding. Second, stopping is delayed rather than eager. Commit decisions do not appear until after Round~3, by which point the graph has already passed through the main repair-heavy stage. The active set then contracts from 512 episodes to 388 after Round~3, 174 after Round~4, and 41 after Round~5, which is consistent with the main-paper reading that later rounds focus on a shrinking hard-case tail rather than uniformly improving all runs.

\subsection{Unrestricted Controller Stress Test}
\label{app:unrestricted_controller}

\paragraph{Setup.}
The submitted runtime constrains the action space by round: round 1 permits support, dependency, contradiction, and skip; rounds 2 and later add evidence and repair; and commitment is disallowed before round 3. The round-wise audit in Table~\ref{tab:appendix_round_action_audit} therefore shows a phased pattern that is partly imposed by the runtime. To test whether the learned critic retains that pattern on its own, we remove the phase-specific action mask, the heuristic fallback, post-hoc calibration, and the runtime safeguards, expose all six role-legal actions in every round, and leave the frozen learned critic as the only scorer. The generator, roles, graph state, maximum rounds, and synthesis are unchanged; we evaluate both the submitted round-3 commit gate and commitment from round 1, retaining only the checks needed to construct executable edits.

\paragraph{Quality.}
Table~\ref{tab:unrestricted_controller} in the main text reports the result. Removing the runtime mechanisms reduces the pooled average from 7.41 to 7.01 with the submitted commit gate, and additionally allowing commitment from round 1 reduces it to 6.96. Because each variant removes four mechanisms at once, we do not attribute that loss to any single one of them.

\paragraph{Behaviour.}
The unrestricted rows of Table~\ref{tab:unrestricted_controller} in the main text show that the learned ordering survives. Dependency and contradiction still dominate round 1 (73.8--75.0\% against 77.5\%), evidence and repair still dominate round 2 (80.5--82.5\% against 83.4\%), and when commitment is allowed from round 1 only 6 of 512 episodes commit before round 3, leaving the mean commit round at 4.16 against 4.18. The learned sequence is therefore not simply an artifact of the inference-time mask or of the commit gate. We nevertheless treat schedule-constrained absences as runtime effects rather than as standalone evidence of learned preference.

\section{Frozen-Snapshot vs Sequential-Update Runtime}
\label{app:frozen_snapshot_runtime}

\paragraph{Setup.}
The main runtime uses frozen snapshots: all active roles in a round observe the same pre-round graph and only the selected role-local decisions are realized afterward. To isolate this runtime choice, we replace it with a sequential-update variant that keeps the graph schema, role set, candidate construction, maximum rounds, final proposal synthesis, and controller family fixed, but executes active roles one by one inside a round so later roles observe earlier same-round mutations. We evaluate two matched comparisons on the same 512-group held-out subset used in the main paper: \textbf{EIG-Sequential-Heuristic} against \textbf{EIG-Heuristic}, and \textbf{EIG-Sequential-DropIn} against \textbf{EIG (Ours)}. For presentation clarity, Table~\ref{tab:ablation_results} collects the sequential rows in the calibration-and-runtime block and their frozen-snapshot anchors in the edits-and-commits and full-system blocks. The sequential rows aggregate over three fixed role orders (canonical, reverse, and cyclic) and report mean $\pm$ standard deviation across those orders.

\paragraph{Results.}
The sequential-update rows of Table~\ref{tab:ablation_results} in the main text show a consistent degradation when frozen-snapshot execution is replaced by sequential same-round updates. Relative to the matched frozen-snapshot anchors, \textbf{EIG-Sequential-Heuristic} drops from 7.05 to 6.80 on AIIB, from 6.72 to 6.64 on LiveIdeaBench, and from 6.89 to 6.72 overall. \textbf{EIG-Sequential-DropIn} drops from 7.69 to 7.16 on AIIB, from 7.12 to 6.78 on LiveIdeaBench, and from 7.41 to 6.97 overall. The benchmark-wise variation across role orders remains modest: the sequential heuristic row shows standard deviations of 0.03 on AIIB and 0.05 on LiveIdeaBench, and the sequential drop-in row shows 0.05 and 0.06, respectively.

\paragraph{Interpretation.}
These results are consistent with the motivation for frozen snapshots: conditioning same-round role-local decisions on a common graph state reduces within-round update dependence and ensures that all role-local proposals address the same graph, making their contributions easier to compare and merge. The comparison should still be read asymmetrically across the two sequential families. \textbf{EIG-Sequential-Heuristic} is a protocol-matched runtime contrast, whereas \textbf{EIG-Sequential-DropIn} reuses the frozen-snapshot-trained two-head critic and its calibration under a different execution semantics. We therefore interpret the drop-in row as a runtime stress test rather than as a fully protocol-matched sequential retraining. Even under that mismatch, the learned-controller sequential variant remains stronger than the sequential heuristic counterpart (6.97 versus 6.72 average), suggesting that learned edit-and-commit control retains value, while the full frozen-snapshot EIG remains strongest overall.

\paragraph{Efficiency implication.}
Beyond quality, the sequential-update variant is substantially slower in wall-clock time. In the frozen-snapshot runtime, all active roles observe the same pre-round graph snapshot and propose edits independently, so their LLM calls can be issued in parallel within a round. The sequential variant executes roles one by one, forcing later roles to wait for earlier generation to complete. With multiple active roles per round, this serial dependency multiplies per-round wall-clock latency by the active-role count. The frozen-snapshot design is therefore not only a quality choice but also a practical efficiency choice for multi-agent collaboration.

\section{Cross-Backbone Sanity Check With Selected Baselines}
\label{app:backbone_sensitivity}

\paragraph{Setup.}
The main paper uses Qwen3-8B as the generator backbone for both EIG and the reproduced baselines so that the primary comparison isolates framework differences rather than backbone mismatch. As an appendix-only portability check, we additionally replace the generator backbone while keeping the benchmark-native evaluator fixed. We evaluate on the same balanced 512-group held-out subset used for the main paper, with 256 AI Idea Bench 2025 groups and 256 LiveIdeaBench groups. Within each backbone block, Direct, AI-Researcher, and EIG use the same generator backbone and the same benchmark-visible input/output contract, so the comparison remains method-centered rather than provider-centered. Scores are reported on the same 0--10 presentation scale used in the main paper, and the AIIB and Live columns show mean $\pm$ standard deviation over 3 seeds. We interpret this appendix experiment as a portability check rather than a controlled model-scaling study.

\input{tables/appendix_backbone_sensitivity_table.tex}

\paragraph{Cross-backbone comparison.}
Table~\ref{tab:appendix_backbone_sanity} should be read horizontally within each backbone block. Under both DeepSeek-V3.2 and GPT-4.1-mini, EIG is the strongest method on AIIB, on LiveIdeaBench, and in the overall average. With DeepSeek-V3.2, EIG reaches 7.64 average, compared with 7.42 for AI-Researcher and 6.74 for Direct. With GPT-4.1-mini, EIG reaches 7.32, compared with 6.73 and 6.05, respectively. The same ranking therefore holds across both backbone substitutions, suggesting that the main-paper gain is not specific to the original Qwen3-8B generator.

\paragraph{Interpretation.}
The appendix result is not meant to claim that all methods scale identically across model families. Instead, it tests whether EIG's advantage survives a backbone swap. The answer is positive in two narrower senses. First, EIG remains ahead of both a one-shot baseline and a strong multi-stage baseline under both alternative backbones. Second, the reported standard deviations remain modest, which suggests that the ranking is stable across the 3 seeds rather than driven by a single favorable run. We therefore interpret these results as an appendix-only portability sanity check showing the same method ranking under two additional backbones.

\section{Human Evaluation Protocol}
\label{app:human_eval_protocol}
\label{app:human_eval}

\subsection{Packet Construction}
\label{app:human_eval_packet}

\paragraph{Sampling.}
The blind human study is designed to evaluate final proposals rather than internal trajectories. We sample 24 held-out paper-eval groups, balanced as 12 AI Idea Bench 2025 groups and 12 LiveIdeaBench groups.

\paragraph{Packet size.}
For each sampled benchmark group, the packet contains one final proposal per compared method under the same benchmark-visible input. This packet size is large enough to expose stable cross-method differences while remaining small enough for careful expert reading by research-trained annotators.

\subsection{Reviewer Pool}
\label{app:human_eval_pool}

\paragraph{Composition.}
We recruit 12 reviewers: seven PhD students, two postdoctoral researchers, two research assistants, and one assistant professor. All reviewers have at least two publications in relevant areas, including AI/ML, NLP and large language models, AI for science, or agentic AI. Table~\ref{tab:human_eval_reviewer_pool} summarizes the reviewer pool by primary area and position; the area labels use each reviewer's primary area only, although several reviewers span more than one category.

\input{tables/human_eval_reviewer_pool_table.tex}

\paragraph{IRB and ethics review.}
The study involved volunteer expert review of anonymized machine-generated text with no identifiable human subjects data. Because reviewers evaluated only anonymized system outputs and no personal or sensitive data were collected, the study was determined to be exempt from full Institutional Review Board review under our institution's policy for minimal-risk research involving expert evaluation of non-identifiable artifacts. All reviewers provided informed consent to participate.

\subsection{Assignment and Blinding}
\label{app:human_eval_blinding}

\paragraph{Blinding.}
For each sampled benchmark group, reviewers receive the benchmark-provided context together with anonymized proposals whose method identities have been removed and whose order has been randomized. Reviewers do not see hidden reference answers, runtime metadata, graph traces, or generation logs.

\paragraph{Assignment rule.}
Each benchmark group is assigned to three reviewers. Assignments are made by matching the group topic to reviewers' primary or secondary area and by requiring self-reported familiarity of at least moderate level whenever possible. With 24 groups and three reviewers per group, the full packet yields 72 group-review assignments, which corresponds to roughly six groups per reviewer on average.

\subsection{Full Instruction Packet and Compensation}
\label{app:human_eval_packet_full}

\paragraph{Recruitment and compensation.}
Reviewers were recruited from the authors' academic network as volunteer domain experts. No monetary compensation was provided. All reviewers consented to participate in the anonymous proposal-ranking study with full knowledge that they were evaluating machine-generated research ideas.

\paragraph{Full instruction text.}
The following verbatim instructions were sent to each reviewer:

\begin{quote}
\textbf{Blind Human Evaluation Instruction Packet}

\textbf{Goal.}
This study evaluates final scientific proposals produced by different systems. We sample 24 held-out paper-evaluation groups, balanced as 12 AI Idea Bench 2025 groups and 12 LiveIdeaBench groups. Reviewers read anonymized proposals and score only the final proposal quality, not the internal reasoning trace or method identity.

\textbf{What reviewers see.}
For each benchmark group, reviewers receive the benchmark-provided context and a shuffled set of anonymized proposals. Method names, runtime metadata, graph traces, scores, and generation logs are hidden. Reviewers should not search for the hidden target paper or use external references beyond the provided context.

\textbf{Rating scale.}
Use a 1--5 integer scale for each criterion. A score of 1 means poor or missing, 3 means acceptable but limited, and 5 means strong. Use the full scale when proposals differ meaningfully. After scoring all proposals in the same benchmark group, give an overall ranking from 1 (best) to 7 (worst), with no ties.

\textbf{Criteria.}
\begin{itemize}
    \item novelty: Does the proposal make a non-trivial research claim relative to the provided context?
    \item significance: Would solving the proposed problem matter scientifically or practically?
    \item feasibility: Is the method concrete enough to execute with plausible data, models, and evaluation?
    \item clarity: Is the proposal coherent, specific, and easy to understand?
    \item context adherence: Does the proposal stay faithful to the benchmark prompt and visible evidence?
    \item overall quality: Considering all criteria together, how strong is the proposal?
    \item overall ranking: After scoring, how would you rank the anonymized proposals within this benchmark group from strongest to weakest?
\end{itemize}

\textbf{Reviewer instructions.}
\begin{enumerate}
    \item Read the benchmark context first.
    \item Read all anonymized proposals for the same benchmark instance before scoring.
    \item Score each proposal independently on all six criteria.
    \item After assigning scores, provide a forced overall ranking of all proposals for the same benchmark group.
    \item Do not reward a proposal for being longer unless the extra detail improves scientific content.
    \item Penalize proposals that invent unsupported context, ignore the benchmark topic, or provide only generic method names without a concrete validation plan.
    \item If two proposals are similar, use feasibility, context adherence, and evaluation specificity to break ties.
\end{enumerate}
\end{quote}

\subsection{Scoring Rubric}
\label{app:human_eval_rubric}

\paragraph{Scoring process.}
Reviewers first read the benchmark context, then read all anonymized proposals for the same group, and finally score each proposal independently on a 1--5 integer scale. The packet instructs reviewers to use the full scale, to avoid rewarding verbosity by itself, and to break close ties by looking more carefully at feasibility, context adherence, and evaluation specificity. Proposals are penalized if they invent unsupported context, drift away from the benchmark topic, or remain generic despite sounding plausible.

\paragraph{Criteria.}
The rubric includes novelty, significance, feasibility, clarity, context adherence, and overall quality. Novelty measures whether the proposal appears meaningfully non-trivial within the provided context; significance measures whether the problem and contribution matter; feasibility measures whether the idea is concrete and executable; clarity measures whether the proposal is coherent and easy to follow; context adherence measures whether the proposal stays faithful to the benchmark prompt and evidence; and overall quality summarizes the complete judgment.

\paragraph{Forced ranking.}
After assigning the six criterion scores, reviewers also provide a forced overall ranking from 1 (best) to 7 (worst) across the anonymized proposals for the same benchmark group. The ranking is used as a complementary preference signal rather than as a replacement for the criterion-level scores.

\subsection{Agreement Summary}
\label{app:human_eval_agreement}

\paragraph{Correlation summary.}
In addition to the mean ratings and mean overall rank reported in the main paper, we summarize reviewer consistency from the three completed annotation packets. On the 24-group packet, the mean pairwise Spearman correlation is 0.62 for the overall-score ordering across methods within a group and 0.54 for the forced within-group ranking itself. Aggregating the three reviewers per group yields a mean Kendall's $W$ of 0.69 over the seven-method rankings, indicating moderate-to-strong agreement on relative preference despite natural variance in absolute scores.

\paragraph{Interpretation.}
The annotation style follows recent human-evaluation practice in scientific ideation studies that rely on experienced research reviewers rather than crowd workers~\citep{si2025novelideas,gu2024scimuse}.

\subsection{Paired Uncertainty}
\label{app:human_eval_uncertainty}

\paragraph{Procedure.}
For each human-evaluation metric, we compare EIG against the strongest non-EIG baseline on that metric using the 72 matched group-review blocks. We report the mean EIG-favoring difference, a 95\% paired bootstrap confidence interval with 20,000 resamples over matched blocks, and a one-sided Wilcoxon signed-rank test in the EIG-favoring direction. For average rank, lower is better, so the reported difference is baseline rank minus EIG rank. Table~\ref{tab:human_eval_paired_uncertainty} reports these statistics, which we use as descriptive paired uncertainty checks rather than as a family-wise multiple-testing claim over all rubric dimensions.

\input{tables/human_eval_paired_uncertainty_table.tex}

\paragraph{Interpretation.}
The paired analysis supports the main human-evaluation claim most strongly on the two summary outcomes: overall quality and average rank both have positive confidence intervals and small Wilcoxon $p$-values against VirSci, the strongest baseline on those metrics. Among the rubric dimensions, the evidence is strongest for significance and clarity. Novelty, feasibility, and context adherence remain positive in mean, but their confidence intervals include zero; we therefore treat them as directional signals rather than conclusive per-dimension wins. This is consistent with the main-paper claim that EIG improves reviewer-perceived proposal quality overall, without requiring every rubric axis to be independently significant.

\section{Case Study}
\label{app:qualitative_case_study}
\label{app:case_study}

\subsection{Case Background}
\label{app:case_background}

\paragraph{Instance.}
We use the instance \texttt{liveideabench-parasites-302}. This case is useful because the strongest text-only draft is already fluent and scientifically plausible. The question is therefore not whether EIG can make a weak draft sound better, but whether the graph state can expose a concrete weakness and route a targeted repair.

\paragraph{Diagnostic focus.}
The matched text-only draft proposes mechanism-aware modeling for host-parasite interactions. Its weakness is specificity: the scientific object remains broad, and the evaluation plan mainly follows a generic robustness template. The EIG trace in Table~\ref{tab:case_study_parasites} shows how the controller converts this weakness into a graph-local repair target and commits a proposal that makes parasite life cycles explicit within a dynamic-graph formulation.

\subsection{Generated Proposal and Graph Evolution}
\label{app:case_trace}

This subsection reports the controller trace and the committed proposal in one boxed cell so that readers can inspect the round-by-round action logic together with the final generated idea.

\input{tables/case_study_parasites_table.tex}

\subsection{Discussion}
\label{app:case_interpretation}

\paragraph{Mechanism-level change.}
The case highlights the type of improvement that EIG is designed to produce. The graph does not simply preserve a longer dialogue history; it keeps the claim, mechanism, risks, and evaluation dependencies inspectable. As a result, the selected actions form a coherent sequence: expose the weak point, ground it, repair it, and commit only after the repaired claim chain is ready for synthesis. The resulting proposal changes the scientific object from broad host-parasite interactions to an explicitly life-cycle-aware dynamic graph model.

\paragraph{Connection to the paper's claim.}
This example complements the aggregate results by showing how a small number of edit-and-commit decisions can affect the final proposal. The main qualitative point is that explicit graph state makes the weak point easier to localize, and the learned critic turns that localization into a concise and interpretable round-by-round improvement path. The direction that path induces is not merely better-organized prose: Table~\ref{tab:case_literature} maps each change it makes onto independent ecological work and the falsifiable test that follows.

\subsection{Literature Consistency of the Induced Direction}
\label{app:case_literature}

\paragraph{What the repair changed.}
The change EIG induces in this case is structural rather than stylistic. Each element the repair rewrites---the modeled object, the mechanism, and the evaluation---corresponds to structure that independent ecological work treats as scientifically substantive, and each carries concrete falsifiable predictions that the matched text-only draft's generic robustness template never specifies.

\input{tables/case_study_literature_table.tex}

\paragraph{What the mapping shows.}
The committed proposal moves from generic interaction prediction to a stage-structured, temporally evolving, heterogeneous mechanism whose assumptions determine specific falsification tests: stage-transition forecasting and host-species holdout for the life-cycle object, temporal extrapolation and typing ablations for the mechanism, and environmental-regime and host-distribution shifts for the evaluation. Because each of these tests follows from the mechanism rather than from a template, the graph repair produces a proposal that is both literature-grounded and directly falsifiable, and the gain is structural and testable rather than a matter of presentation. The supporting work spans both the foundational treatments of these structures and recent independent results from 2024--2025, so the induced direction is consistent with how the field currently treats the problem rather than only with its older literature.

\section{Limitations}
\label{app:limitations}

\paragraph{Scope.}
We study the ideation stage: whether a system produces stronger research proposals under benchmark-visible inputs. Whether an idea then survives implementation is a separate question, since an experimental outcome also depends on data, tuning, compute, and domain expertise downstream of the proposal. Isolating EIG's contribution there needs its own protocol, which we leave to future work.

\paragraph{Supervision.}
The controller is trained from protocol-matched heuristic labels rather than expert action annotation. Those labels are relative preferences within a validated candidate slate, and the same-state audit shows the learned policy departs from the teacher on 37.7\% of edit decisions while improving the pooled average by 0.52. Human-, LLM-, or outcome-grounded teachers are a natural next step.

\paragraph{Representation.}
The node, relation, and action schema is deliberately compact, which is what lets a single controller transfer across two benchmarks and three backbones without per-domain tuning. Richer structures, such as nested hypotheses or competing evaluation frameworks, would need a larger action space, and we leave them to future work.

%% file: tables/notation_table.tex
\begin{table}[ht]
\centering
\tablefont
\renewcommand{\arraystretch}{0.98}
\caption{Key notation used in parallel EIG.}
\label{tab:notation}
\begin{tabular*}{\textwidth}{@{\extracolsep{\fill}}p{0.24\textwidth} p{0.68\textwidth}}
\toprule
\textbf{Symbol} & \textbf{Meaning} \\
\midrule
$x$ & Benchmark-visible ideation input. \\
$\mathcal{L}(x)$ & Benchmark-permitted literature or reference context. \\
$y$ & Final structured proposal. \\
$G_t=(V_t,E_t)$ & Typed idea graph at round $t$. \\
$\bar G_t$ & Frozen graph snapshot used inside round $t$. \\
$H_t$ & Cached shared-encoder representation of $G_t$. \\
$\mathcal{M}$ & Set of role-specialized agents. \\
$\mathcal{M}_t$ & Active roles at round $t$. \\
$\mathcal{S}_t^{(r)}$ & Validated role-local candidate slate for role $r$. \\
$d_t^{(r)}$ & Selected role decision for role $r$ at round $t$. \\
$\Delta_t^{(r)}$ & Role-local patch materialized relative to the pre-round snapshot. \\
$\Delta_t$ & Deterministically merged non-empty patch set. \\
$T(G_t,\Delta_t)$ & Realized graph transition after patch merge. \\
$q_{\theta}^{\mathrm{edit}}$ & Edit head for role-local action selection. \\
$q_{\theta}^{\mathrm{commit}}$ & Commit head for post-round stopping. \\
$\mathcal{J}$ & Joint training objective for edit and commit supervision. \\
$\tau$ & Commit round. \\
$G_{\tau}$ & Committed graph used for final proposal synthesis. \\
$\mathcal{E}(G_{\tau})$ & Evidence attached to the committed proposal subgraph. \\
$p_{\psi}$ & Final proposal synthesizer conditioned on graph, evidence, and input. \\
\bottomrule
\end{tabular*}
\end{table}

%% file: tables/runtime_objects_table.tex
\begin{table*}[t]
\centering
\tablefont
\setlength{\tabcolsep}{5pt}
\renewcommand{\arraystretch}{0.98}
\caption{Persistent runtime objects used by the EIG implementation.}
\label{tab:runtime_objects}
\begin{tabular*}{\textwidth}{@{\extracolsep{\fill}}p{0.15\textwidth} p{0.46\textwidth} p{0.29\textwidth}}
\toprule
\textbf{Object} & \textbf{Key fields} & \textbf{Role in runtime / training} \\
\midrule
Node & identifier, type, text, generating role, branch identifier, confidence score, evidence list, active/inactive status, timestamp, provenance & {\raggedright Carries proposal state edited by roles and encoded by the critic.\par} \\
Edge & source and target node identifiers, relation type, role, branch, optional evidence identifier, note, resolution flag, timestamp & {\raggedright Records typed support, contradiction, dependency, and repair relations.\par} \\
Branch & node identifiers, edge identifiers, frozen flag, rejected flag, notes & {\raggedright Tracks one role-local trajectory before or after consolidation.\par} \\
Graph action & round, role, action kind, target identifiers, payload, rationale, source, timestamp & {\raggedright Stores replayable decisions and training supervision rows.\par} \\
\bottomrule
\end{tabular*}
\end{table*}

%% file: tables/appendix_teacher_student_table.tex
\begin{table}[t]
\centering
\tablefont
\setlength{\tabcolsep}{6pt}
\renewcommand{\arraystretch}{0.98}
\caption{Teacher-student audit on the same held-out subset as Table~\ref{tab:ablation_results}. The submitted signal heuristic is re-scored counterfactually on the exact logged learned graphs and candidate slates, so both policies decide on identical states and differ only in the decision rule. \textbf{EIG} is the final submitted policy, including learned scoring, calibration, diversity checks, heuristic fallback, and runtime safeguards. \textbf{Difference} is EIG minus the heuristic, with a paired 95\% bootstrap interval over the matched groups; positive values favor EIG.}
\label{tab:teacher_student}
\setlength{\tabcolsep}{14.0pt}
\begin{tabular}{lcccccc}
\toprule
\rowcolor{headertint}
\textbf{Benchmark} & \textbf{Groups} & \textbf{EIG-Heuristic} & \textbf{EIG (Ours)} & \textbf{Difference} & \textbf{95\% CI} \\
\midrule
AIIB & 256 & 7.05 & \textbf{7.69} & +0.64 & [0.24, 0.65] \\
Live & 256 & 6.72 & \textbf{7.12} & +0.40 & [0.25, 0.45] \\
Pooled & 512 & 6.89 & \textbf{7.41} & +0.52 & [0.21, 0.54] \\
\bottomrule
\end{tabular}
\end{table}

%% file: tables/appendix_teacher_student_agreement_table.tex
\begin{table}[t]
\centering
\tablefont
\setlength{\tabcolsep}{6pt}
\renewcommand{\arraystretch}{0.98}
\caption{Decision agreement between the submitted heuristic and the final EIG policy on identical audited states, from the same audit as Table~\ref{tab:teacher_student}. \textbf{Agreement} is the share of decisions on which the two policies select the same action. Edit decisions are role-local action choices; commit decisions are post-round graph-level stopping decisions.}
\label{tab:teacher_student_agreement}
\setlength{\tabcolsep}{34.8pt}
\begin{tabular}{lccc}
\toprule
\rowcolor{headertint}
\textbf{Decision} & \textbf{Same} & \textbf{Different} & \textbf{Agreement} \\
\midrule
Edit & 4,216 & 2,552 & 62.3\% \\
Commit & 1,607 & 271 & 85.6\% \\
\bottomrule
\end{tabular}
\end{table}

%% file: tables/training_signal_profile_table.tex
\begin{table*}[t]
\centering
\tablefont
\setlength{\tabcolsep}{5pt}
\renewcommand{\arraystretch}{0.98}
\caption{Offline supervision corpus and profiling overhead for the two-head EIG graph critic. Token counts are traced generation tokens in the profiling corpus; no benchmark scoring calls are included. We report tokens rather than currency because provider prices and discounts are not fixed in the artifact.}
\label{tab:training_signal_profile}
\begin{tabular*}{\textwidth}{@{\extracolsep{\fill}}p{0.32\textwidth} p{0.58\textwidth}}
\toprule
\textbf{Item} & \textbf{Value} \\
\midrule
Profiled benchmark groups & 400 total: 200 AI Idea Bench 2025 and 200 LiveIdeaBench \\
Train/dev groups & 300 train groups (150/150 by benchmark) and 100 dev groups (50/50 by benchmark) \\
Edit supervision & 5,327 pre-action states and 63,833 candidate rows; 48,004 train and 15,829 dev candidate examples \\
Commit supervision & 1,725 post-round states: 400 commit and 1,325 continue; 1,294 train and 431 dev examples \\
Profiling overhead & 5.55M traced tokens in total, 13.88k tokens per profiled run on average, and 1.04k tokens per transition on average \\
Held-out paper evaluation & 512 disjoint groups: 256 AI Idea Bench 2025 and 256 LiveIdeaBench; overlap with graph-critic train/dev is zero \\
\bottomrule
\end{tabular*}
\end{table*}

%% file: tables/qwen_v2_inference_cost_table.tex
\begin{table}[t]
\centering
\tablefont
\renewcommand{\arraystretch}{0.98}
\caption{Supplementary inference-time cost analysis on the available Qwen3-8B paper-evaluation batch summaries. Tokens are averaged over 512 held-out paper-eval samples, and all rows are as-submitted configurations; the budget-matched budgets used for the main comparison in Table~\ref{tab:main_results} are reported in Table~\ref{tab:matched_budget}. We report traced generation tokens only; benchmark scoring calls and the offline graph-critic profiling cost in Table~\ref{tab:training_signal_profile} are excluded. Lower is better.}
\label{tab:qwen_v2_inference_cost}
\setlength{\tabcolsep}{7.6pt}
\begin{tabular}{lcccc}
\toprule
\rowcolor{headertint}
\textbf{Method} & \textbf{Prompt Tokens} & \textbf{Completion Tokens} & \textbf{Total} & \textbf{$\times$ Direct} \\
\midrule
Direct & 1376 & 309 & 1685 & 1.00 \\
Self-Refine\venue{NeurIPS'23}~\citep{madaan2023selfrefine} & 5013 & 997 & 6010 & 3.57 \\
Graph of Thoughts\venue{AAAI'24}~\citep{besta2024graphthoughts} & 14302 & 3490 & 17792 & 10.56 \\
AI-Researcher\venue{ICLR'25}~\citep{si2025novelideas} & 8908 & 2216 & 11124 & 6.60 \\
SciPIP\venue{arXiv'24}~\citep{wang2024scipip} & 5062 & 1446 & 6508 & 3.86 \\
VirSci\venue{ACL'25}~\citep{su2025many} & 5329 & 811 & 6140 & 3.64 \\
\rowcolor{eighighlight}
\textbf{EIG (Ours)} & 12229 & 1522 & 13750 & 8.16 \\
\bottomrule
\end{tabular}
\end{table}

%% file: tables/appendix_matched_budget_table.tex
\begin{table}[t]
\centering
\tablefont
\setlength{\tabcolsep}{6pt}
\renewcommand{\arraystretch}{0.98}
\caption{Inference-budget matching on the fixed 512-group held-out subset used in Table~\ref{tab:main_results}. The upper block reports each multi-stage baseline as submitted; the middle block reports the same method after method-native scaling, which raises its traced budget to at or above EIG's 13.75k; the lower block is the unscaled EIG reference. \textbf{Tokens (K)} is the average traced generation budget per held-out sample. Error bars are sample standard deviations over three runs; higher is better for \textbf{AIIB}, \textbf{Live}, and \textbf{Avg.}.}
\label{tab:matched_budget}
\setlength{\tabcolsep}{8.3pt}
\begin{tabular}{llcccc}
\toprule
\rowcolor{headertint}
\textbf{Configuration} & \textbf{Method} & \textbf{AIIB} & \textbf{Live} & \textbf{Avg.} & \textbf{Tokens (K)} \\
\midrule
\multirow{3}{*}{As submitted} & AI-Researcher\venue{ICLR'25}~\citep{si2025novelideas} & 7.09{\scriptsize$\pm$0.09} & 6.78{\scriptsize$\pm$0.08} & 6.94 & 11.12 \\
 & SciPIP\venue{arXiv'24}~\citep{wang2024scipip} & 6.93{\scriptsize$\pm$0.09} & 6.78{\scriptsize$\pm$0.08} & 6.86 & 6.51 \\
 & VirSci\venue{ACL'25}~\citep{su2025many} & 6.97{\scriptsize$\pm$0.08} & 6.79{\scriptsize$\pm$0.07} & 6.88 & 6.14 \\
\midrule
\multirow{3}{*}{Budget-matched} & AI-Researcher\venue{ICLR'25}~\citep{si2025novelideas} & 7.10{\scriptsize$\pm$0.02} & 6.80{\scriptsize$\pm$0.02} & 6.95 & 14.81 \\
 & SciPIP\venue{arXiv'24}~\citep{wang2024scipip} & 6.94{\scriptsize$\pm$0.02} & 6.79{\scriptsize$\pm$0.03} & 6.87 & 16.42 \\
 & VirSci\venue{ACL'25}~\citep{su2025many} & 6.95{\scriptsize$\pm$0.02} & 6.81{\scriptsize$\pm$0.02} & 6.88 & 14.14 \\
\midrule
\rowcolor{eighighlight}
Reference & \textbf{EIG (Ours)} & \textbf{7.69}{\scriptsize$\pm$0.09} & \textbf{7.12}{\scriptsize$\pm$0.08} & \textbf{7.41} & 13.75 \\
\bottomrule
\end{tabular}
\end{table}

%% file: tables/appendix_round_action_audit_table.tex
\begin{table*}[t]
\centering
\tablefont
\setlength{\tabcolsep}{3.6pt}
\renewcommand{\arraystretch}{0.98}
\caption{Round-wise controller action audit on the 512-group held-out EIG evaluation subset. For Support, Evidence, Repair, Dependency, Contradiction, and Skip, percentages are normalized by the round-wise selected role-action total from the controller trace export. Commit is a post-round graph-level decision and is therefore reported as a raw count only.}
\label{tab:appendix_round_action_audit}
\tablefontdense
\setlength{\tabcolsep}{8.0pt}
\begin{tabular}{lccccccc}
\toprule
\rowcolor{headertint}
\textbf{Round} & \textbf{Support} & \textbf{Evidence} & \textbf{Repair} & \textbf{Dependency} & \textbf{Contradiction} & \textbf{Skip} & \textbf{Commit} \\
\midrule
1 & 576 (22.5\%) & 0 (0.0\%) & 0 (0.0\%) & 960 (37.5\%) & 1024 (40.0\%) & 0 (0.0\%) & 0 \\
2 & 365 (14.3\%) & 984 (38.4\%) & 1152 (45.0\%) & 24 (0.9\%) & 35 (1.4\%) & 0 (0.0\%) & 0 \\
3 & 256 (10.0\%) & 816 (31.9\%) & 950 (37.1\%) & 298 (11.6\%) & 0 (0.0\%) & 240 (9.4\%) & 124 \\
4 & 0 (0.0\%) & 652 (33.6\%) & 570 (29.4\%) & 188 (9.7\%) & 0 (0.0\%) & 530 (27.3\%) & 214 \\
5 & 0 (0.0\%) & 230 (26.4\%) & 240 (27.6\%) & 32 (3.7\%) & 0 (0.0\%) & 368 (42.3\%) & 133 \\
6 & 0 (0.0\%) & 16 (7.8\%) & 23 (11.2\%) & 13 (6.3\%) & 0 (0.0\%) & 153 (74.6\%) & 41 \\
\bottomrule
\end{tabular}
\end{table*}

%% file: tables/appendix_backbone_sensitivity_table.tex
\begin{table*}[t]
\centering
\tablefont
\setlength{\tabcolsep}{7pt}
\renewcommand{\arraystretch}{0.98}
\caption{Appendix-only cross-backbone sanity check on a balanced 512-group held-out subset. Within each backbone block, all methods use the same generator backbone and the same benchmark-native evaluator; only the ideation framework differs. Scores are benchmark-native results rescaled to 0--10 for presentation; AIIB and Live report mean $\pm$ standard deviation over 3 seeds, and Avg.\ reports the corresponding overall mean. This appendix experiment is intended as a portability check rather than a controlled scaling study.}
\label{tab:appendix_backbone_sanity}
\setlength{\tabcolsep}{14.0pt}
\begin{tabular}{llccc}
\toprule
\rowcolor{headertint}
\textbf{Backbone} & \textbf{Method} & \textbf{AIIB} & \textbf{Live} & \textbf{Avg.} \\
\midrule
\multirow{3}{*}{DeepSeek-V3.2}
& Direct & 6.53{\scriptsize$\pm$0.11} & 6.94{\scriptsize$\pm$0.09} & 6.74 \\
& AI-Researcher\venue{ICLR'25}~\citep{si2025novelideas} & 7.66{\scriptsize$\pm$0.09} & 7.18{\scriptsize$\pm$0.11} & 7.42 \\
\rowcolor{eighighlight}
& \textbf{EIG (Ours)} & \textbf{7.97}{\scriptsize$\pm$0.08} & \textbf{7.31}{\scriptsize$\pm$0.10} & \textbf{7.64} \\
\midrule
\multirow{3}{*}{GPT-4.1-mini}
& Direct & 5.71{\scriptsize$\pm$0.11} & 6.39{\scriptsize$\pm$0.08} & 6.05 \\
& AI-Researcher\venue{ICLR'25}~\citep{si2025novelideas} & 6.88{\scriptsize$\pm$0.10} & 6.58{\scriptsize$\pm$0.09} & 6.73 \\
\rowcolor{eighighlight}
& \textbf{EIG (Ours)} & \textbf{7.58}{\scriptsize$\pm$0.11} & \textbf{7.08}{\scriptsize$\pm$0.08} & \textbf{7.33} \\
\bottomrule
\end{tabular}
\end{table*}

%% file: tables/human_eval_reviewer_pool_table.tex
\begin{table*}[t]
\centering
\tablefont
\renewcommand{\arraystretch}{0.98}
\setlength{\tabcolsep}{6pt}
\caption{Reviewer composition for the blind human study. Counts use each reviewer's primary area only; several reviewers have secondary experience in adjacent areas. Res.\ Asst.\ denotes research assistant, and Asst.\ Prof.\ assistant professor.}
\label{tab:human_eval_reviewer_pool}
\setlength{\tabcolsep}{15.6pt}
\begin{tabular}{lccccc}
\toprule
\rowcolor{headertint}
\textbf{Area} & \textbf{PhD} & \textbf{Postdoc} & \textbf{Res.\ Asst.} & \textbf{Asst.\ Prof.} & \textbf{Total} \\
\midrule
AI/ML & 2 & 0 & 1 & 0 & 3 \\
NLP / LLM & 2 & 1 & 0 & 0 & 3 \\
AI4Sci & 2 & 0 & 1 & 1 & 4 \\
Agentic AI / Automation & 1 & 1 & 0 & 0 & 2 \\
\midrule
Total & 7 & 2 & 2 & 1 & 12 \\
\bottomrule
\end{tabular}
\end{table*}

%% file: tables/human_eval_paired_uncertainty_table.tex
\begin{table*}[t]
\centering
\tablefont
\setlength{\tabcolsep}{4.0pt}
\renewcommand{\arraystretch}{0.98}
\caption{Paired human-evaluation uncertainty against the strongest baseline for each metric. Positive differences favor EIG; for average rank, this means the baseline has a larger worse rank.}
\label{tab:human_eval_paired_uncertainty}
\setlength{\tabcolsep}{5.4pt}
\begin{tabular}{llccccc}
\toprule
\rowcolor{headertint}
\textbf{Metric} & \textbf{Strongest baseline} & \textbf{EIG} & \textbf{Baseline} & \textbf{Diff.} & \textbf{95\% CI} & $\boldsymbol{p}$ \\
\midrule
Novelty & Graph of Thoughts\venue{AAAI'24}~\citep{besta2024graphthoughts} & \textbf{3.04} & 2.94 & +0.10 & [-0.10, 0.29] & 0.162 \\
Significance & VirSci\venue{ACL'25}~\citep{su2025many} & \textbf{3.78} & 3.14 & +0.64 & [0.38, 0.90] & $<10^{-4}$ \\
Feasibility & VirSci\venue{ACL'25}~\citep{su2025many} & \textbf{3.88} & 3.71 & +0.17 & [-0.13, 0.46] & 0.114 \\
Clarity & VirSci\venue{ACL'25}~\citep{su2025many} & \textbf{3.74} & 3.28 & +0.46 & [0.11, 0.79] & 0.014 \\
Context & SciPIP\venue{arXiv'24}~\citep{wang2024scipip} & \textbf{3.15} & 3.08 & +0.07 & [-0.17, 0.31] & 0.305 \\
Overall & VirSci\venue{ACL'25}~\citep{su2025many} & \textbf{4.01} & 3.50 & +0.51 & [0.36, 0.65] & $<10^{-4}$ \\
Avg. Rank & VirSci\venue{ACL'25}~\citep{su2025many} & \textbf{1.64} & 2.92 & +1.28 & [0.88, 1.68] & $<10^{-4}$ \\
\bottomrule
\end{tabular}
\end{table*}

%% file: tables/case_study_parasites_table.tex
\begin{table*}[htbp]
\centering
\tablefont
\setlength{\tabcolsep}{0pt}
\setlength{\fboxsep}{7pt}
\caption{Case study on \texttt{liveideabench-parasites-302}. The boxed cell shows the vertical round trace and the final committed proposal in one place.}
\label{tab:case_study_parasites}
\begin{tabular}{@{}c@{}}
{\setlength{\fboxsep}{7pt}\setlength{\fboxrule}{0.4pt}%
\fcolorbox{black!35}{gray!8}{%
\begin{minipage}{\dimexpr0.98\textwidth-2\fboxsep-2\fboxrule\relax}
\textbf{Round-by-round controller trace}

The strongest text-only draft is fluent but broad: it proposes generic host-parasite interaction modeling without a sufficiently specific mechanism-domain link. EIG therefore uses a short sequence of intentional graph edits before commit.

\smallskip
\textbf{Round 1. Structural support}

\textbf{Actions.} Add support edges that connect the problem, mechanism, and evaluation branches into a readable claim chain.

\textbf{Selection rationale.} Because the draft is already coherent at the sentence level, the controller first improves graph inspectability rather than rewriting content immediately.

\textbf{Graph effect.} The graph now makes the central weakness explicit: the mechanism remains too generic and is only weakly tied to parasite-specific biological structure.

\smallskip
\textbf{Round 2. Evidence attachment}

\textbf{Actions.} Attach evidence to the mechanism-side node from the mechanism-aware modeling branch.

\textbf{Selection rationale.} Before proposing a repair, the controller grounds the candidate mechanism so that the next edit is justified by graph-local evidence rather than by a generic theme.

\textbf{Graph effect.} The mechanism branch becomes better supported, which sharpens the diagnosis that the missing element is explicit parasite life-cycle structure.

\smallskip
\textbf{Round 3. Targeted repair}

\textbf{Actions.} Propose repair edits that inject parasite life-cycle knowledge and a temporal graph mechanism into the proposal state.

\textbf{Selection rationale.} The critic identifies weak coupling between the existing mechanism and the parasite domain, so the selected edit is the one that most directly repairs this mismatch.

\textbf{Graph effect.} The proposal shifts from broad host-parasite interaction modeling toward a mechanism-aware dynamic graph formulation that explicitly models parasite life-cycle structure.

\smallskip
\textbf{Round 4. Completion and commit}

\textbf{Actions.} Attach final grounding evidence to the remaining assumption node and commit the graph.

\textbf{Selection rationale.} Once the repaired claim chain is stable, further edits have low marginal value; the better decision is to stop and synthesize.

\textbf{Graph effect.} The final proposal is generated from a graph whose problem, mechanism, risk, and evaluation nodes are already aligned.

\bigskip
\textbf{Final generated idea proposal}

\emph{Mechanism-Aware Modeling of Parasite-Host Dynamics with Dynamic Graphs}

\textbf{Problem.} Current predictive models for host-parasite networks typically treat interactions as static or model all entities as homogeneous nodes. This ignores that parasites have stage-structured life cycles requiring specific hosts or conditions to progress, and that interaction outcomes depend on dynamic host states and temporally varying environmental drivers. Consequently, these models struggle to generalize to unseen host species, novel environmental regimes, or perturbations such as climate shifts.

\textbf{Core hypothesis.} Explicitly encoding parasite life-cycle stage transitions and host-state dynamics as a temporally evolving heterogeneous graph improves predictive performance on out-of-distribution host-parasite interactions compared to static or homogeneous graph models.

\textbf{Method.} We represent host-parasite systems as temporal heterogeneous graphs where node types distinguish host entities from parasite life-cycle stages, and edges are typed by interaction mode (infection, transmission, stage progression). The mechanism-aware component constrains message passing: transition functions between parasite stages are parameterized separately from host-host or host-parasite edges, reflecting biological state dependencies. A temporal graph neural network processes time-ordered graph snapshots, conditioning predictions on topological history and available time-varying node features (e.g., host traits, environmental covariates, immune readouts) through modular encoders.

\textbf{Evaluation.} Core tasks are temporal link prediction and state-transition forecasting. Out-of-distribution generalization is tested via (i) temporal extrapolation, (ii) host-species holdout, and (iii) environmental regime shift. We report AUROC/AUPRC for link prediction, macro-F1 for state transitions, and Expected Calibration Error for reliability. Ablations isolate the contribution of heterogeneous typing, explicit temporal modeling, and life-cycle transition constraints against appropriate baselines. Learned dynamics are validated against biologically documented case studies with known parasite life cycles.

\bigskip
\textbf{Case takeaway}

The qualitative gain is not mainly stylistic. The selected actions expose a weak mechanism-domain link, ground that diagnosis with evidence, repair it with a life-cycle-aware graph mechanism, and commit only after the repaired claim chain is ready for synthesis.
\end{minipage}}}
\\
\end{tabular}
\end{table*}

%% file: tables/case_study_literature_table.tex
\begin{table*}[t]
\centering
\setlength{\tabcolsep}{6pt}
\renewcommand{\arraystretch}{0.98}
\caption{Literature consistency of the direction the graph repair induces in \texttt{liveideabench-parasites-302}. Each element the repair changes---the modeled object, the mechanism, and the evaluation---corresponds to structure that independent ecological work treats as scientifically substantive, and each implies a specific falsifiable test that the matched text-only draft's generic robustness template does not specify. \textbf{Change} is what EIG alters relative to the text-only draft, \textbf{Independent literature} is the supporting work, and \textbf{Implied test} is the resulting falsifiable prediction.}
\label{tab:case_literature}
\begin{tabular*}{\textwidth}{@{\extracolsep{\fill}}p{0.22\textwidth} p{0.38\textwidth} p{0.27\textwidth}}
\toprule
\textbf{Change} & \textbf{Independent literature} & \textbf{Implied test} \\
\midrule
Modeled object: parasite life-cycle stages and dynamic host states, replacing broad host--parasite interactions & Parker et al. (2003, Nature) treat complex life cycles as ordered movement through successive hosts, with stage- and host-specific survival and transmission consequences; Silva and Koella (2025, PLOS Pathog.) show that interactions within a life cycle shape virulence evolution & Stage-transition forecasting; host-species holdout \\
\addlinespace
Mechanism: typed temporal transitions in a heterogeneous graph, replacing generic mechanism-aware modeling & Poisot et al. (2015, Oikos) argue that ecological interactions vary through space and time; Pilosof et al. (2017, Nat. Ecol. Evol.) motivate multilayer networks for multiple interaction types; Yin and Rudolf (2024, Ecol. Lett.) give a framework for temporal community structure; Dom\'inguez-Garc\'ia and K\'efi (2024, PLOS Comput. Biol.) analyse the structure and robustness of networks with two interaction types & Temporal extrapolation; ablations of node and edge typing and of life-cycle constraints \\
\addlinespace
Evaluation: temporal extrapolation, host-species holdout, environmental-regime shift, and mechanism ablation, replacing generic robustness checks & Cable et al. (2017, Phil. Trans. R. Soc. B) show that climatic and anthropogenic change alters parasite transmission through habitat, biodiversity, host demography, and evolution; Theodosopoulos et al. (2025, Glob. Change Biol.) report a climate-driven increase in parasite transmission over the last quarter century & Environmental-regime and host-distribution stress tests \\
\bottomrule
\end{tabular*}
\end{table*}